# Strain-induced Dynamic Spin-Phonon Coupling in Epitaxial $RuO_2$ Films

In Hyeok Choi[1,†], Seung Gyo Jeong[2,3†], Jae Hyuck Lee[4], San Kang[4], Sreejith Nair[2], Changyoung Kim[4], Dirk Wulferding[5], Bharat Jalan[*,2], Jong Seok Lee[*,1]

[1]*Department of Physics and Photon Science, Gwangju Institute of Science and Technology (GIST), Republic of Korea*

[2] *Department of Chemical Engineering and Materials Science, University of Minnesota−Twin Cities, Minneapolis, Minnesota 55455, United States*

[3]*Department of Physics, Hankuk University of Foreign Studies, Yongin 17035, Republic of Korea*

[4]*Department of Physics and Astronomy, Seoul National University, Seoul 08826, Republic of Korea*

[5]*Department of Physics and Astronomy, Sejong University, Seoul 05006, Republic of Korea*

**Summary**

Magnetic order parameters can couple to quantized lattice vibration via both piezomagnetic and magnetoelastic effects, leading to the renormalization of phonon dispersion. Here, we demonstrate photo-induced dynamic frequency modulation of THz phonons excited in anisotropically-strained epitaxial $RuO_2$ thin films using ultrafast coherent phonon spectroscopy and time-resolved magneto-optic Kerr effect measurement. A coherent oscillation of a transverse acoustic phonon appears in the sub-THz range with increasing film thickness above 4 nm due to local dislocation arising from the anisotropic strain relaxation, which hosts large non-zero shear strain. Remarkably, this mode exhibits a pronounced temperature-dependent transient hardening below approximately 500 K. While this behavior does not by itself establish spin-phonon coupling, its temperature evolution is consistent with the characteristic temperature range previously associated with magnetic transition in strained $RuO_2$, suggesting that magnetic correlations may contribute to the observed phonon renormalization. Furthermore, an optical phonon oscillation emerges in the transient polarization rotation of the photo-induced non-equilibrium state, significantly softened near the critical temperature, while there is no observable transient polarization rotation in fully-strain-relaxed films. Such notable dynamic frequency modulations in acoustic and optical phonons offer an opportunity to manipulate phonons in the THz range through the spin-phonon coupling controlled by epitaxial design, which can inspire a new class of ultrafast quantum opto-spintronics.

[†]Equally contributed

[*]Corresponding authors: bjalan@umn.edu, jsl@gist.ac.kr

Altermagnets exhibit non-relativistic spin splitting in the electronic band structure without net magnetization, classified as *d*-wave, *g*-wave, or *i*-wave magnets following the coupled crystal symmetry[1-4]. This unusual behavior is not caused by relativistic effects like spin-orbit coupling but instead arises from the specific symmetry of the crystal structure. In these materials, crystal symmetry relates two sublattices with opposite spin orientations via a rotational operation, resulting in time-reversal symmetry breaking without net magnetization[1-4].

Owing to the strong coupling between magnetic order and lattice structure, the control of the crystalline lattice and its vibrational degree of freedom in such magnetic materials provides a unique avenue to manipulate magnetic orders. Specifically, strain can couple to staggered magnetization through two types of interactions: piezomagnetic (linear coupling) and magnetoelastic (quadratic coupling). This means that applying mechanical stress (or strain)[5] can change the magnetic state, enabling a control of magnetic domains. Beyond these static effects, dynamic interactions between spins and the lattice can also play a critical role. For example, staggered magnetization can modulate the sound velocity or frequency of phonons propagating through the material[6-9]. Furthermore, dynamic fluctuations of magnetic order can couple to phonons of specific symmetries[10], even in multi-domain states, leading to a direct coupling between magnons and phonons. As a result, observing changes in phonon behavior can provide an indirect, yet powerful way to detect the presence of spin-phonon coupling in crystal-symmetry-coupled magnets.

Rutile $RuO_2$ has been reported to be either *d*-wave magnetic or also non-magnetic, as demonstrated by various studies[11-20]. These conflicting results suggest the important role of extrinsic factors, such as defects, material quality, or strain. In particular, recent studies using $RuO_2$ (110) epitaxial films have demonstrated that large strain from the substrate can significantly modify the electronic band structure near the Fermi level, resulting in a stain-induced magnetic state[17,21-25]. Indeed, experimental evidence for the strain-induced magnetic state in epitaxial $RuO_2$ (110) films has been provided by optical second-harmonic generation (SHG) and anomalous Hall measurements[4,17]. This suggests that the *d*-wave magnetic state can be realized in $RuO_2$ through strain engineering, indicating a strong coupling between the crystal lattice and the magnetic order[1].

In this work, we demonstrate frequency engineering of THz coherent phonons in systematically controlled $RuO_2$ epitaxial films. We employed ultrafast coherent phonon spectroscopy and time-resolved magneto-optic Kerr effect (MOKE) measurements (as illustrated in Fig. 1) to probe the phonon dynamics in $RuO_2$ thin films grown on $TiO_2$ (110) substrates using hybrid molecular beam epitaxy (MBE). When $RuO_2$ was excited with ultrafast laser pulses, we observed coherent oscillations corresponding to acoustic or optical phonons. These oscillations exhibit clearly temperature-dependent anomalies in their frequencies near $T_c$~500 K, suggesting the existence of spin-phonon coupling. This onset temperature is close to the transition temperature observed in our previous SHG measurements in fully-strained $RuO_2$ films[4]. Notably, even in partially relaxed (but still anisotropically strained) $RuO_2$ films, temperature-dependent SHG signal variations, absent in bulk $RuO_2$, were observed near similar temperatures[4]. One possible explanation is that local strain fields originating from dislocations associated with strain relaxation induce the observed anomalies. Our results reveal that under photo-excitation, THz phonon dynamics can be modulated in ultrafast time-scale in anisotropically strained epitaxial $RuO_2$ films. These reveal the temperature-dependent evolution of these phonon modes under nonequilibrium conditions and highlight their sensitivity to lattice-spin coupling in epitaxial $RuO_2$. Furthermore, these findings point to new opportunities for ultrafast manipulation of lattice and spin dynamics through epitaxial strain.

Figure 2a displays two representative results of transient reflectivity ($\Delta R$) of the 12 nm $RuO_2$ (110) thin film measured at room temperature. $\Delta R$ was obtained as a function of the pump-probe time delay ($t$) with a sampling rate of 5 THz. Here, the sampling rate is defined as the inverse of the time step used in the measurement. Both pump and probe pulses are centered at a wavelength of 785 nm. The pump polarization was set along $[1\bar{1}0]$ direction, while the probe polarization was aligned along either the [001] ($\varphi$ = 0°) and $[1\bar{1}0]$ ($\varphi$ = 90°) direction, as schematically illustrated in the top panel of Fig. 2a. The two $\Delta R(t)$ curves exhibit a notable difference in their maximum amplitudes, which is attributed to the anisotropic electronic structure of $RuO_2$ (110) film[21,26]. After subtracting the exponentially decaying background (dashed lines) to obtain the residual oscillation component $\Delta R_{oscill}(t)$, we found that $\Delta R_{oscill}(t)$ exhibits well-defined oscillations (inset of Fig. 2a). We observed both fast sub-THz and slower GHz oscillations, which exhibit different dependencies on probe polarization and the film thickness. The GHz oscillation frequency remains identical with an increase in the film thickness, as shown in Figs. 2b and 2c, and its amplitude $A_{oscill}$ for 12 nm $RuO_2$ film is independent of $\varphi$, as shown in Fig. 2d (Section S1 in Supporting Information for more detail). These results imply that the GHz oscillation originates from the longitudinal acoustic (LA) sound wave in the substrate due to interference between reflected beams from the film surface and the strain wave propagating into the substrate[27] (Section S2, Supporting Information). On the other hand, the sub-THz oscillation exhibits a systematic thickness-dependence of oscillation frequency (Figs. 2b and 2c), suggesting that it arises from standing acoustic waves confined within the film, rather than from bulk optical phonon modes[28-31]. Furthermore, in Fig. 2d, the $A_{oscill}$ of the sub-THz oscillation has a two-fold anisotropy, with a maximum at $\varphi$ = 0°, indicating that the oscillation corresponds to a transverse acoustic (TA) mode.

Figure 2e presents the anisotropic strain relaxation of the 12 nm $RuO_2/TiO_2$ epitaxial film along the in-plane [001] direction where the sub-THz oscillation is maximized (Section S3, Supporting Information). The left and right panels of Fig. 2e show the XRD reciprocal space mapping (RSM) results for the (310) and (332) Bragg reflections, respectively. Along the in-plane reciprocal lattice direction $[1\bar{1}0]$ ($Q_{1\bar{1}0}$), the peak position of the $RuO_2$ film aligns with that of the $TiO_2$ substrate, indicating that the film is fully strained along this direction. On the other hand, along the [001] in-plane direction ($Q_{001}$), the peak position of $RuO_2$ deviates from that of $TiO_2$ but does not reach the bulk $RuO_2$ value ($Q_{001} \sim 0.646$ Å$^{-1}$), implying partial strain relaxation. These results demonstrate an anisotropically-strained state of 12 nm $RuO_2/TiO_2$ epitaxial film, fully strained along $[1\bar{1}0]$ and partially relaxed along [001], as schematically shown in the right of Fig. 2d. This strain relaxation along the [001] direction can accompany edge dislocations[4,32], where abrupt changes in the lattice constant introduce large shear strain $\eta$, particularly the $\eta_{xz}$ component. Here, the $x$, $y$, and $z$ coordinates correspond to the [001], $[1\bar{1}0]$, and [110] crystallographic directions, respectively. Upon photo-excitation, the presence of non-zero $\eta_{xz}$ can drive the generation of TA phonon modes[30,31]. The absence of TA coherent oscillations in fully strained films strongly supports a close relationship between edge dislocations and TA phonons. The dynamic modulation of shear strain $\Delta\eta_{xz}$ can be detected via $R(t)$ due to the photo-induced modulation of the linear dielectric tensor components[33], following the relation $\Delta\varepsilon_{xx} \sim -\varepsilon_{xx}^2\Delta\eta_{xz}$, leading to the observed maximum $A_{\text{oscill}}$ at $\varphi = 0°$, as shown in Fig. 2d. Assuming that the coherent oscillations are generated only in the strain-relaxed regions of the films, a standing wave would form an antinode mode. The corresponding sound velocity $v_s$ can then be estimated using the standing wave equation $v_s = 2f\cdot(d - d_0)$, where $f$ is the frequency of the standing wave, $d$ is the thickness of the film, and $d_0$ is the averaged thickness of strained area[4]. The sound velocity estimated from the oscillation frequency (~6000 m/s) is in good agreement with the theoretically calculated TA mode velocity in $RuO_2$[34], providing additional evidence of photo-induced TA phonon generation in the $RuO_2$ thin films (Section S4, Supporting Information).

The photo-excited TA mode for $RuO_2$ (110) films exhibits time-varying mode softening or hardening depending on temperature $T$, suggesting dynamic modulation of elastic properties. The top panel of Fig. 3a presents a contour plot of $\Delta R_{\mathrm{oscill}}$ for the 12 nm $RuO_2$ (110) film as functions of time and $T$. The middle and bottom panels of Fig. 3a show $\Delta R_{\mathrm{oscill}}(t)$ at two selected $T$, 560 K and 100 K, respectively. Scattered symbols in the top panel denote the local maxima (blue) and minima (orange), obtained via quadratic function fitting (Section S5, Supporting Information). Vertical dashed lines in the middle and bottom panels indicate the expected peak positions assuming a constant oscillation period (i.e., $P_{01} = P_{12} = P_{23}$). At 560 K (the middle panel of Fig. 3a), all peaks (red circles) appear on the right side of dashed lines, and the difference between the positions of peaks and lines become larger as time increases, indicating TA mode softening. In contrast, at 100 K (the bottom panel of Fig. 3a), the peaks (blue circles) shift to the left of the dashed lines with time, reflecting TA mode hardening. To quantitatively analyze the peak shift, we defined a $t$-dependent quasi-oscillation frequency, $f_{nm} = 0.5/P_{nm}$ (with $n$, $m$ = 0, 1, 2, and 3), and summarized the results in Fig. 3b. The extracted $f_{nm}$ at both 100 K and 560 K clearly shows $t$-dependence, with changes of up to about 10%. Figure 3c further exhibits the $T$-dependent evolution of $f_{01}$, $f_{12}$ and $f_{23}$ across a broader $T$ range. We note that the error bars are calculated using the least squares method[35]. As $T$ decreases, the sign of their difference $f_{01} - f_{12}$ and $f_{01} - f_{23}$ changes from positive to negative at ~420 K (inset of Fig. 3c), indicating a transition from dynamic mode hardening to softening. While similar time-varying mode frequencies upon photoexcitation have been reported in other materials[36-38], they are typically attributed to electron-phonon decoupling[37,38], the carrier diffusion[36], or the modification of ion potentials[38]. However, these mechanisms generally explain either softening or hardening individually. A temperature-driven crossover between the two regimes, as observed in the $RuO_2$ (110) film, has not been previously reported.

Notably, all $f_{nm}$ parameters exhibit a clear onset behavior around $T_c$ ~500 K, similar with the $T$ at which the previous optical SHG study reported the $T$-dependent anomaly associated with the magnetic transition for strained $RuO_2/TiO_2$ (110) film[4]. Furthermore, it is noteworthy that the $RuO_2$ thin films do not show any structural phase transition across $T_c$[4]. This suggests that the strong $T$-dependent anomalies in the phonon frequency are likely related to the spin-phonon coupling; the TA mode in $RuO_2$ can be renormalized through piezomagnetic or magnetoelastic coupling. The distinct piezomagnetic response in magnets with 4/*mmm* point group enables momentum-selective coupling between the magnetic octupole and the TA mode involving lattice vibrations within the (001) plane with $B_{1g}$ symmetry[3], whereas conventional magnetoelastic coupling is not directionally constrained. In our experiments, however, the observed TA mode originates from [001] strain relaxation of $RuO_2$ on $TiO_2$ (110) substrates and propagates along the [110] direction. This specific geometry constrains the accessible phonon propagation directions and therefore limits the direct applicability of the piezomagnetic coupling concept to the present measurements. Instead, we attribute the observed behavior to magnetoelastic coupling, which can also modify the acoustic phonon band[6-9] with less sensitivity to the phonon propagation direction. Since the photo-excited TA mode relaxes within ~4 ps, we consider $f_{23}$ at $t$ = 3.4 ps as representative of the equilibrium state frequency. As shown in Fig. 3c, this quasi-static TA mode frequency decreases by about 10% as $T$ increases from 10 K to $T_c$, a change comparable to magnetoelastic phonon renormalization observed in other magnetic oxides[8,9]. In addition, we observed slope changes in $f_{01}$ and $f_{23}$ upon heating around 250 K. This behavior is reminiscent of anomalies in X-ray scattering intensity and electrical resistivity previously reported at the similar temperature in $RuO_2$[39,40], suggesting possible changes in electronic, spin, or phononic correlations; however, a more detailed investigation of the low-temperature regime is beyond the scope of the present study.

This TA phonon softening via magnetoelastic coupling provides a consistent explanation for the time-varying mode frequency following the photo-excitation**.** Figure 3d illustrates the photo-induced renormalization of the TA mode on the ultrafast timescale. In the quasi-equilibrium state, the TA mode exhibits the phonon softening as the temperature approaches $T_c$ (top panel of Fig. 3d). Upon the photo-excitation, the electron temperature ($T_e$) rapidly increases on the femtosecond timescale, and the electron system thermalizes with the lattice within a few picoseconds, comparable to the decay time of coherent oscillation. In addition, the spin temperature ($T_s$) is considered to immediately follow $T_e$, as electron-spin thermalization typically occurs within sub-picoseconds in metallic transition metal oxides[41,42]. In the magnetic state ($T < T_c$), the initial photo-excited increase in $T_s$ ($\approx T_e$) leads to the transient mode softening, with the phonon frequency gradually recovering to its initial value during cooling (bottom panel of Fig. 3d). Based on the two-temperature model (Section S6, Supporting Information), a photo-induced elevation in $T_e$ from 297 K to 380 K yields an estimated frequency shift of approximately –0.04 THz from the quasi-static value $f_{23}$ (Fig. 3c), in good agreement with $f_{01}$–$f_{23}$ values at $T = 297$ K (inset of Fig. 3c). Above $T_c$, on the other hand, $f_{01}$–$f_{23}$ still exhibits a non-zero positive value, suggesting the possible existence of photo-induced electron-phonon decoupling[37]. We note that the structural and thermal effect upon lattice heating can be ruled out due to their negligible contributions (Section S7, Supplementary

Information). Furthermore, we confirmed that the time-varying mode exhibits a clear pump-power dependence, supporting a spin–phonon coupling origin for the TA frequency softening or hardening (Section S8, Supplementary Information). This self-consistency supports the interpretation that the time-varying mode hardening below $T_c$ arises from strong TA mode softening near $T_c$ due to magnetoelastic coupling.

Importantly, the significant phonon softening near $T_c$ is revealed not only in the TA mode but also in the $B_{1g}$ optical phonon mode. $B_{1g}$ optical phonons have been reported to couple to the antiferromagnetic order in other rutile systems[43]. To detect the lowest frequency of $B_{1g}$ optical phonons theoretically expected at about 5 THz[44], we increased the temporal resolution of the measurements of $\Delta R(t)$ and transient Kerr rotation $\Delta\theta_K(t)$, corresponding to a maximum detectable spectral width of about 10 THz (See Methods). We note that the ultrafast pump-probe approach selectively excites and probes the dynamics of coherent optical phonons in the $RuO_2$ layer, avoiding the substrate contribution that is often strongly observed in conventional static Raman measurements, and thus provides a more reliable observation to study phonon dynamics in ultrathin $RuO_2/TiO_2$ heterostructures. Figures 4a and 4b display representative results of $\Delta R(t)$ and $\Delta\theta_K(t)$, respectively, obtained with probe polarization aligned along $\varphi = 0°$ at 300 K under an external magnetic field of 120 mT by alternating its direction. By summing and subtracting the $\Delta\theta_K(t)$ obtained under opposite magnetic field directions, we can extract the field-independent and field-dependent $\Delta\theta_K(t)$, respectively. Notably, the field-independent signals would be attributed to artifact contribution from imperfect balancing or the dynamics of an antiferromagnetic order parameter since it is barely changed under the external magnetic field. On the other hand, the field-dependent signals reflect the dynamics of ferromagnetic order parameter induced by the external magnetic field (Section S9, Supporting Information).

Figure 4b displays the field-independent $\Delta\theta_K(t)$ (black line) with the oscillatory components $\Delta\theta_{K,oscill}(t)$ (blue line). In the 20 THz of sampling rates, while $\Delta R(t)$ shows only the sub-THz oscillation with no observable $B_{1g}$ mode (blue line in Fig. 4a), $\Delta\theta_K(t)$ exhibits an obvious signature of a faster THz oscillation during its rise (blue line in Fig. 4b). We note that the theoretical expectation for non-zero $\Delta\theta_K(t)$ response has been previously reported in the magnetically ordered state of $RuO_2$[45]. Figure 4c displays the Fast Fourier Transform (FFT) spectrum of $\Delta\theta_{K,oscill}(t)$, revealing a broad peak with the center frequency at 4.2 THz. A peak with similar energy appears in the Raman spectrum of the $RuO_2$ single crystal (Fig. 4d), which is assigned to the $B_{1g}$ mode according to the selection rule (Section S10, Supporting Information). This suggests that the observed THz peak in $\Delta\theta_{K,oscill}(t)$ is related to the $B_{1g}$ phonon mode. Moreover, the field-dependent $\Delta\theta_{K,oscill}(t)$ also exhibits similar fast oscillation with a comparable frequency, further supporting its identification as a magnetically coupled mode (Section S9, Supporting Information). Consequently, our observations suggest that the $B_{1g}$ optical phonon, strongly coupled with magnetization dynamics, is both generated and detected through the Kerr response in the anisotropically strained $RuO_2$ film.

The coherent optical phonon oscillation can be generated and observed in $\Delta R$ via displacive excitation of coherent phonons (DECP) and impulsive stimulated Raman scattering (ISRS) process[46]. However, the optical excitation of antisymmetric $B_{1g}$ mode cannot be explained by the DECP mechanism, as it only excites fully symmetric phonon modes. Furthermore, for the 4/*mmm* point group of $RuO_2$ (110), the Raman tensor of the $B_{1g}$ mode contains only off-diagonal components, thus forbidding coherent excitation via the ISRS mechanism, which requires nonzero diagonal elements in the Raman tensor. This is consistent with the absence of $B_{1g}$ mode in the FFT spectrum of $\Delta R_{oscill}(t)$ in Fig. 4c. Similarly, the $B_{1g}$ mode observed in $\Delta\theta_{K,oscill}(t)$ cannot be directly induced by such mechanisms. Instead, we consider that the generation and detection of the $B_{1g}$ coherent oscillation occur during the thermalization process in the non-equilibrium spin system, where strong perturbation of both spin and lattice degrees of freedom allows such coupling. We observed the $B_{1g}$ mode for both φ = 0° and 90°, supporting a detection mechanism distinct from ISRS (Section S11, Supporting Information). The broad spectral width of $B_{1g}$ mode in $\Delta\theta_{K,oscill}(t)$, compared to that in the Raman spectrum (Fig. 4d), indicates a short phonon lifetime of about 0.4 ps, corresponding to a 2.3 THz bandwidth. This suggests that the $B_{1g}$ mode, as being closely coupled to the coherent magnetization dynamics, only transiently exists in the non-equilibrium state following photoexcitation.

Figure 4e displays the $T$-dependent $\Delta\theta_{K,oscill}(t)$, showing a notable change in its oscillation frequency as a function of $T$. FFT analysis (Fig. 4f) and the extracted center frequency (Fig. 4g) shows the broad frequency softening where the frequency is minimized at ~450 K that is slightly lower than $T_c$. This frequency softening of $B_{1g}$ mode near $T_c$ can be elucidated by its coupling with magnetic order[10,43] (Fig. 4h). Interestingly, the oscillation amplitude does not show a significant $T$-dependent variation. This might be owing to the short-range magnetic correlations persisting above the nominal $T_c$ in thin films[47,48], or the magnetic field applied during the measurement although the in-depth understanding requires further investigations. It is noteworthy that we cannot observe any $T$-dependent anomaly in $B_{1g}$ modes in $RuO_2$ single crystals (Section S12, Supporting Information), demonstrating that the photo-induced magnetization dynamics is unique for the anisotropically-strained $RuO_2$ films. We further note that the (110) strain can stabilize magnetism in $RuO_2$ films[17]. Although strain in our system is partially relaxed, we expect that large local strain fields arising from dislocations or strain discontinuities, as well as atomic-scale distortions induced by the optically pumped phonons, can amplify magnetic behavior. Furthermore, we cannot observe any notable anomaly in both magneto-optic signals and oscillations for fully-strain-relaxed 40 nm $RuO_2$ film, which further supports the important role of local and anisotropic strain induced by edge dislocations in the magnetization dynamics in $RuO_2$ (Section S13, Supporting Information).

In summary, we experimentally demonstrated the ultrafast engineering of THz phonons in anisotropically-strained epitaxial $RuO_2$ thin film. By deliberately controlling the film thickness, we induced anisotropic strain relaxation, allowing the generation of TA phonon oscillations in the sub-THz frequency domain. The TA phonon oscillations exhibit a pronounced $T$-dependent frequency change below $T_c$ (~500 K), indicative of possible spin-phonon coupling and resulting thermally-driven time-varying mode hardening. Furthermore, we observed $B_{1g}$ phonon mode at ~5 THz in the transient polarization rotation, which also shows a prominent mode softening near $T_c$. The $T$-dependent anomalies in both acoustic and optical modes reveal a manifestation of spin-phonon coupling in $RuO_2$ thin films. While our measurements do not directly resolve the underlying magnetic order, these results highlight the interplay between lattice dynamics and spin degrees of freedom in epitaxial $RuO_2$. Our findings also open new avenues for THz phononic applications for ultrafast engineering of the elasticity in crystal-symmetry-coupled magnets through the epitaxial heterostructure.

# Method

## Hybrid molecular beam epitaxy

Crystallographic-orientation-controlled epitaxial $RuO_2$ films were grown on $TiO_2$ (110) and (100) single crystalline substrates (*Crystec*) using a hybrid oxide molecular beam epitaxy (MBE) system (*Scienta Omicron*). The hybrid MBE technique with metal-organic precursor has reported exceptional quality of $RuO_2$ thin films with a record-low residual resistivity and an atomically flat surface, enabling precise and systematic analysis down to a nanometer scale[4,49,50]. We utilized the metal-organic precursor $Ru(acac)_3$, which was thermally evaporated from a low-temperature effusion cell (*MBE Komponenten*) between 170-180 °C. The $TiO_2$ substrate underwent a treatment process involving acetone, methanol, and isopropanol, followed by 2-hour baking at 200 °C in a load lock chamber. Subsequently, a 20-minute annealing step was performed in oxygen plasma at 300 °C before the film growth. The oxygen plasma was generated at 250 W and an oxygen gas pressure of $5 \times 10^{-6}$ Torr. After the growth, the sample was cooled to 120 °C in the presence of oxygen plasma to prevent the potential formation of oxygen vacancies. We monitored the film surfaces before, during, and after growth using in-situ reflection high-energy electron diffraction (RHEED) from Staib Instruments. We confirmed the roughness, crystallinity, and film thickness using atomic force microscopy, high-resolution XRD, and scanning transmission electron microscopy measurements[4,49]. We performed the reciprocal space mapping (RSM) using a Rigaku SmartLab XE to examine the strain state.

## Single crystal synthesis

Single crystals of $RuO_2$ were synthesized by sintering a pelletized $RuO_2$ powder inside an alumina crucible in air. The pellet was heated up to 1473 K and slowly cooled down to 1023 K at a cooling rate of 1.25 K per hour until the furnace was turned off for quenching. The blue-black metallic crystals were at maximum 1×1×0.5 $mm^3$ in size.

**Optical pump-probe measurement**

To explore the dynamics of phonons in the photo-induced non-equilibrium states, we monitored transient reflectivity changes using the pump-probe method. Using the sharp-edge long-pass filter and short-pass filter (*Semrock*), we separated the laser beam that has 785 nm center wavelengths and 80 MHz repetition rate (Vision-S, *Coherent*) into pump and probe beam. At this pump energy, free carriers and Ru 4d-$t_{2g}$ electrons are excited. From the interferometric autocorrelation of the probe pulse, we determined an optimized pulse width of approximately 50 fs, corresponding to a maximum spectral width of up to about 10 THz. The optical penetration depth of $RuO_2$ is approximately 26 nm[4], and hence we monitored the response of the entire film with a thickness smaller than 19 nm. Both pump and probe beams are focused on the sample with a beam size of 20 μm in the normal incidence. We set the power of the pump and probe beam as 50 mW and 3 mW, respectively. The pump and probe beam were modulated by an electro-optic modulator with 10 MHz and a mechanical chopper with 700 Hz, respectively. To determine the zero-time delay between pump and probe beams, we measured the transient reflectivity of a semiconductor reference sample (GaAs), which exhibits a step-like rise within our temporal resolution. The data were then fitted using an exponential rise function convoluted with a Gaussian pulse to determine the zero-time delay. We monitored the pump-induced reflectivity changes which are demodulated by side-band frequencies composed of 10 MHz and 700 Hz using the two-channel digital Lock-in amplifier (*Zurich instrument*). To obtain Kerr rotation signals, we modulated a probe beam by photo-elastic modulator (PEM) with 100 kHz. By demodulating detector signals by side-band frequencies composed of 10 MHz and 100 kHz, we obtained pump-induced Kerr signals. All time-resolved MOKE experiments were conducted with the magnetic field applied. We note that the direction of magnetic-field was alternatively changed for each measured point to extract field-dependent and field-independent signals. For the temperature-dependent measurements, $RuO_2$ thin films were attached to the heating stage (*Linkam*) with silver paste in the ambient atmospheric condition and also to the cryogenic system for the low-temperature measurements.

## Acknowledgements

Authors acknowledge valuable discussions with Charles R. W. Steward, Rafael M. Fernandes, and Jörg Schmalian. J.S.L. and I.H.C. acknowledge the support from the Korea government (MSIT) No. RS-2024-00486846. The work by J.H.L., and C.K. was supported by Global Research Development Center (GRDC) Cooperative Hub Program through the National Research Foundation of Korea (NRF) funded by the Ministry of Science and ICT(MSIT) (Grant No. RS-2023-00258359) and the NRF grant funded by MSIT (Grant No. NRF-2022R1A3B1077234). Film synthesis (S.G.J and B.J.) was supported by the U.S. Department of Energy through grant Nos. DE-SC0020211, and (partly) DE-SC0024710. Structural characterization (at UMN) were supported by the Air Force Office of Scientific Research (AFOSR) through Grant Nos. FA9550-21-1-0025, and FA9550-24-1-0169. S.N. was supported partially by the UMN MRSEC program under Award No. DMR-2011401. Parts of this work were carried out at the Characterization Facility, University of Minnesota, which receives partial support from the NSF through the MRSEC program under Award No. DMR-2011401. D.W. acknowledges support from the Institute of Applied Physics of Seoul National University, from the ITRC Program through the IITP, and from the Global Research Development Center (GRDC) Cooperative Hub Program through the NRF, funded by the (MSIT) (Grant Nos. RS-2024-00437191 and RS-2023-00258359).

## Contributions

I.H.C. and S.G.J. contributed equally to this work. J.S.L. and B. J. supervised the project. I.H.C. and J.S.L. performed the pump-probe experiments and analyzed the data. S.G.J., S.N., and B.J. prepared thin films and characterized them. J.H.L. and C.K. prepared single crystals and strain-relaxed films and characterized them. J.H.L. and D.W. conducted Raman spectroscopy experiments. I.H.C., S.G.J., B.J. and J.S.L. wrote the manuscript. All the authors discussed the results, implications, and commented on the manuscript.

## Competing Interests

The authors declare no competing interests.

## Data Availability

All data supporting the results within this paper and supporting information are available from the corresponding author upon reasonable request. Source data are provided in this paper.

## Code Availability

All numerical simulation codes employed in this work are available from the corresponding author upon reasonable request.

# Figures

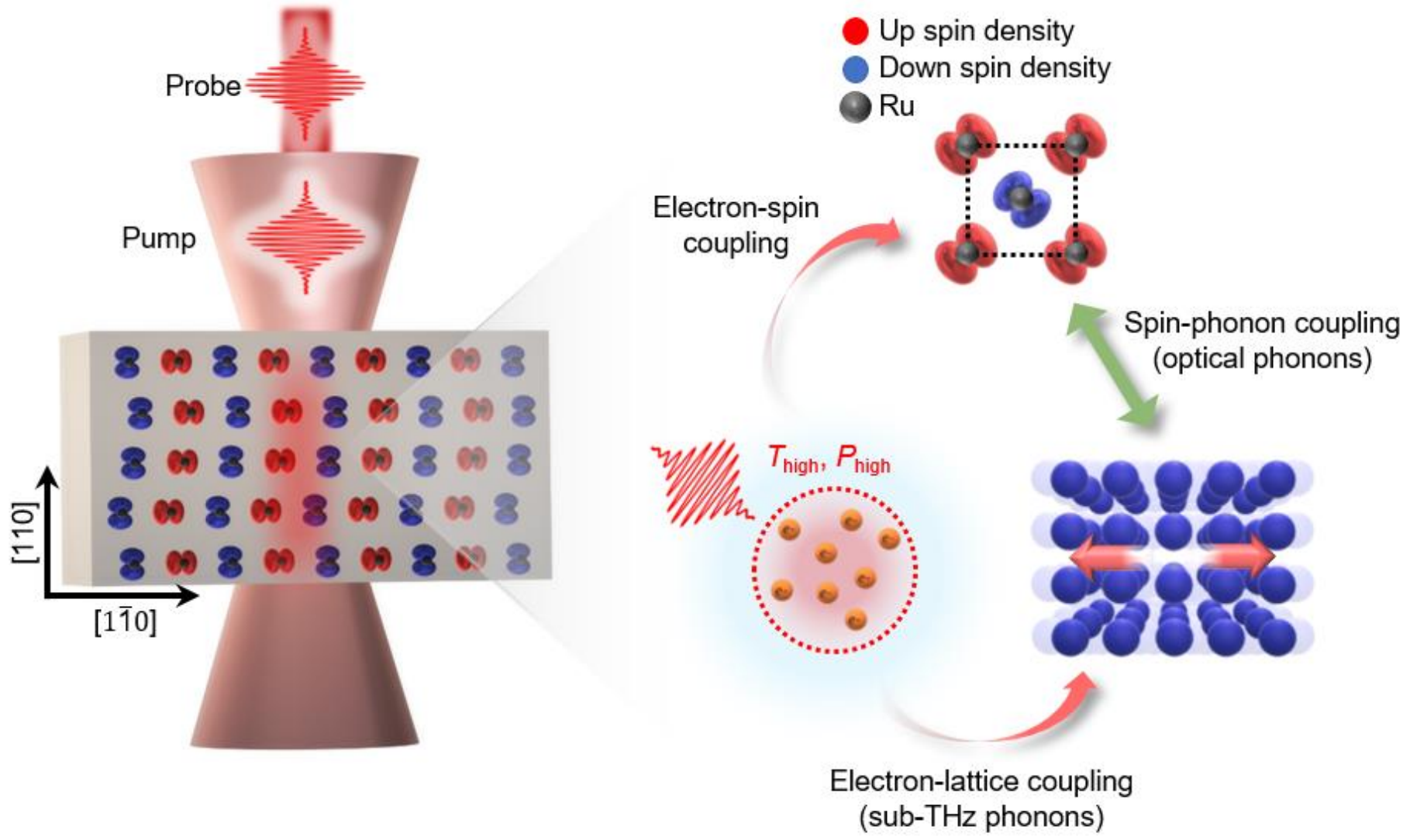

**Figure 1. Dynamic spin-phonon coupling in $RuO_2$ films.** (Left panel) A combination of crystal rotational [$C_{4z}$] and time-reversal [$C_2$] symmetry leads to a coupling between the magnetic order parameter and phonons results in the renormalization of the phononic system. To monitor the phonon dynamics, we measured the pump-induced transient reflectivity ($\Delta R$) and transient magneto-optic Kerr angle ($\Delta\theta_K$), revealing the coherent phonon oscillations and their couplings with the magnetism in $RuO_2$ films. (Right panel) Energy flow diagram following photo-excitation in $RuO_2$ thin films. Upon photo-excitation, photons are solely absorbed by the electron system, leading to an increase in its temperature. Through electron–spin and electron–lattice couplings, energy is subsequently transferred to the spin and lattice systems, respectively. while the spin and lattice systems interact with each other. This dynamic coupling among the three systems leads to the excitation of both acoustic and optical phonons.

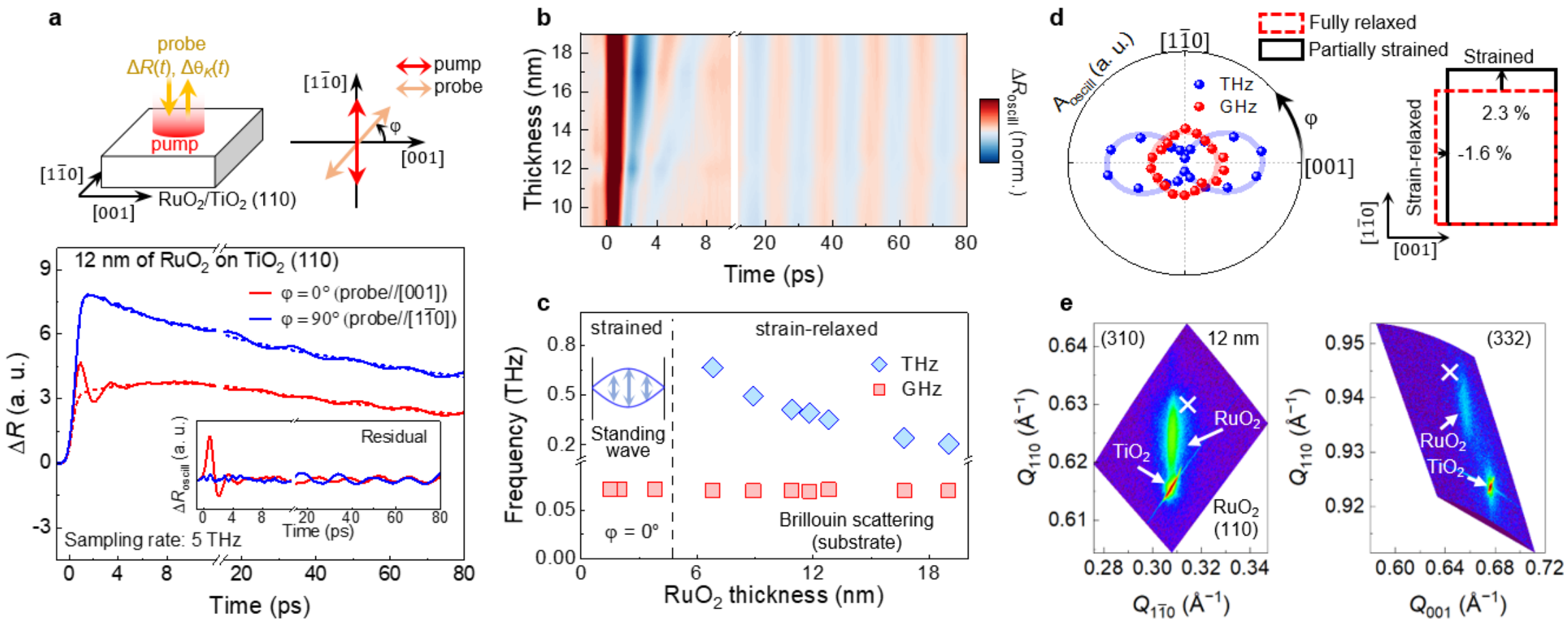

**Figure 2. Observation of transverse acoustic standing wave in $RuO_2$ (110) films. (a)** Pump-induced Δ*R* obtained in 12 nm $RuO_2$ (110) film at two different probe polarization angles φ = 0° and 90° at room temperature. A THz oscillation emerges only when the probe polarization is along the [001] direction (φ = 0°). By fitting the curves with a two-exponential rise-decay model, we extracted pure oscillation parts $\Delta R_{oscill}$ (inset). **(b)** Contour plots of normalized coherent oscillation components $\Delta R_{oscill}$ obtained at room temperature as a function of film thicknesses and time delay when the probe polarization is along the [001] direction. The coherent oscillations are normalized with respect to the maximum peak amplitude $A_{oscill}$. **(c)** Film-thickness-dependent oscillation center frequencies obtained from FFT spectra obtained at φ = 0°. The THz oscillation frequency is systematically hardened as the film thickness decreases, suggesting the THz oscillations are the acoustic standing wave confined in the film, while there is no significant thickness dependence in GHz oscillations. **(d)** φ-dependent $A_{oscill}$ in 12 nm $RuO_2$ (110) film obtained at room temperature. **(e)** Anisotropic strain relaxation observed in XRD reciprocal space mapping (RSM) results of 12 nm $RuO_2$ (110) film around $TiO_2$ (310) and (332) planes. The strain is relaxed only along the [001] direction, where the oscillation amplitude is the maximum.

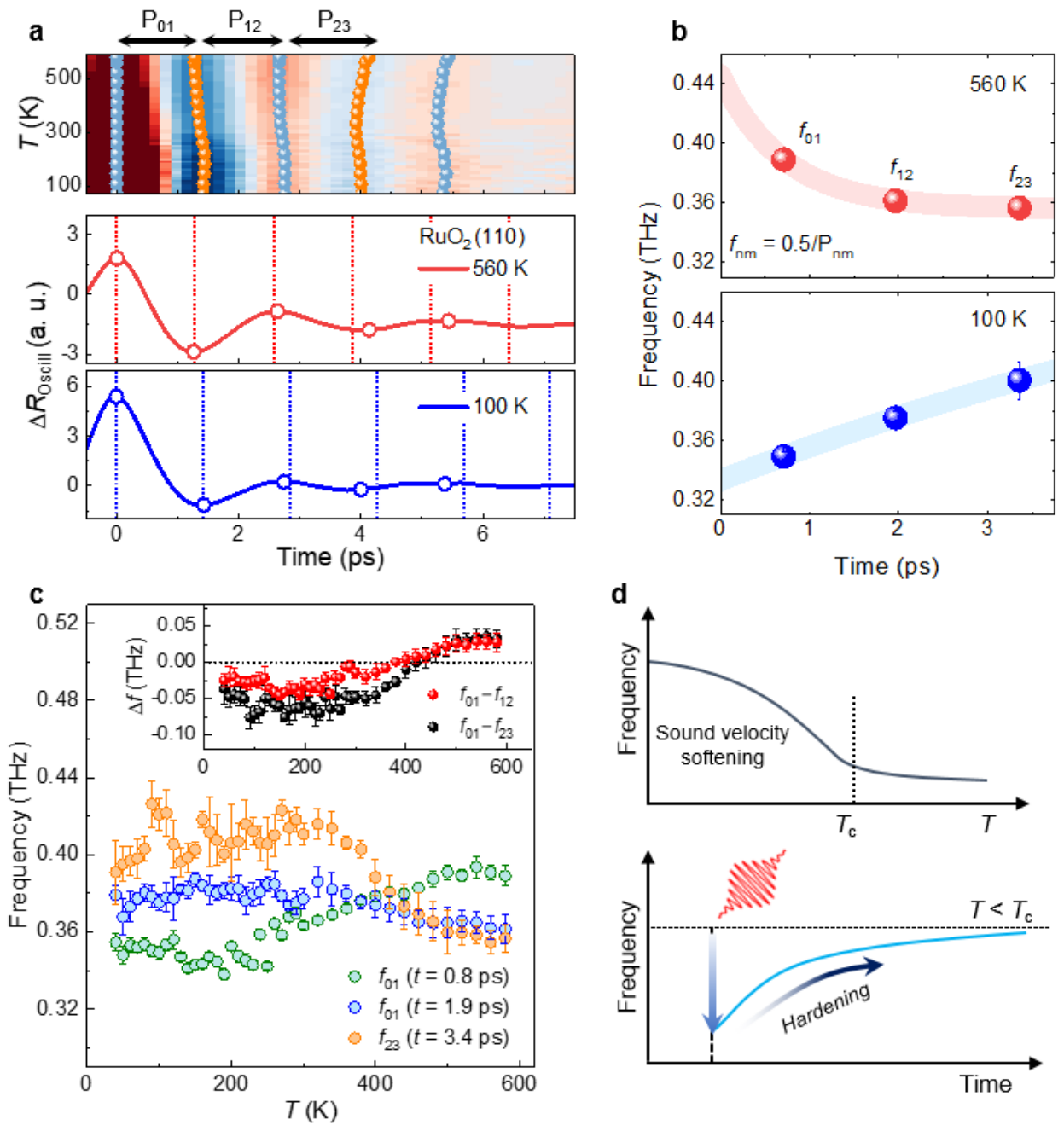

**Figure 3. Temperature-dependent changes in the dynamics of coherent oscillation in 12 nm $RuO_2$ (110) film.** **(a)** Time-dependent coherent oscillations as a function of temperature ($T$) (top panels) and representative oscillations at 100 K and 560 K (middle and bottom panels). Dot symbols represent the maximum and minimum position of oscillation obtained by fitting with the quadratic function. $P_{nm}$ ($n$, $m$ =0, 1, and 2) are defined as the time interval between maximum and minimum peaks for the given $T$. The vertical dashed lines indicate the estimated position by assuming $P_{nm}$ constant to $P_{01}$. **(b)** Time-evolution of $f_{nm}$ at 560 K (upper panel) and 100 K (lower panel), and **(c)** $T$-dependent frequencies $f_{nm}$ for 12 nm $RuO_2$ (110) film. The frequencies $f_{nm}$ are calculated from $P_{nm}$ ($f_{nm}$ = $0.5/P_{nm}$). The inset shows the frequency difference $\Delta f$ between $f_{01}$ and $f_{12}$ ($f_{01}-f_{12}$) and also between $f_{01}$ and $f_{23}$ ($f_{01}-f_{23}$) as a function of $T$. The positive value of frequency difference indicates the frequency hardening with increasing time while the negative value of that indicates the frequency softening. **(d)** Schematic of the possible origin for time-dependent mode hardening below the magnetic phase transition temperature $T_c$.

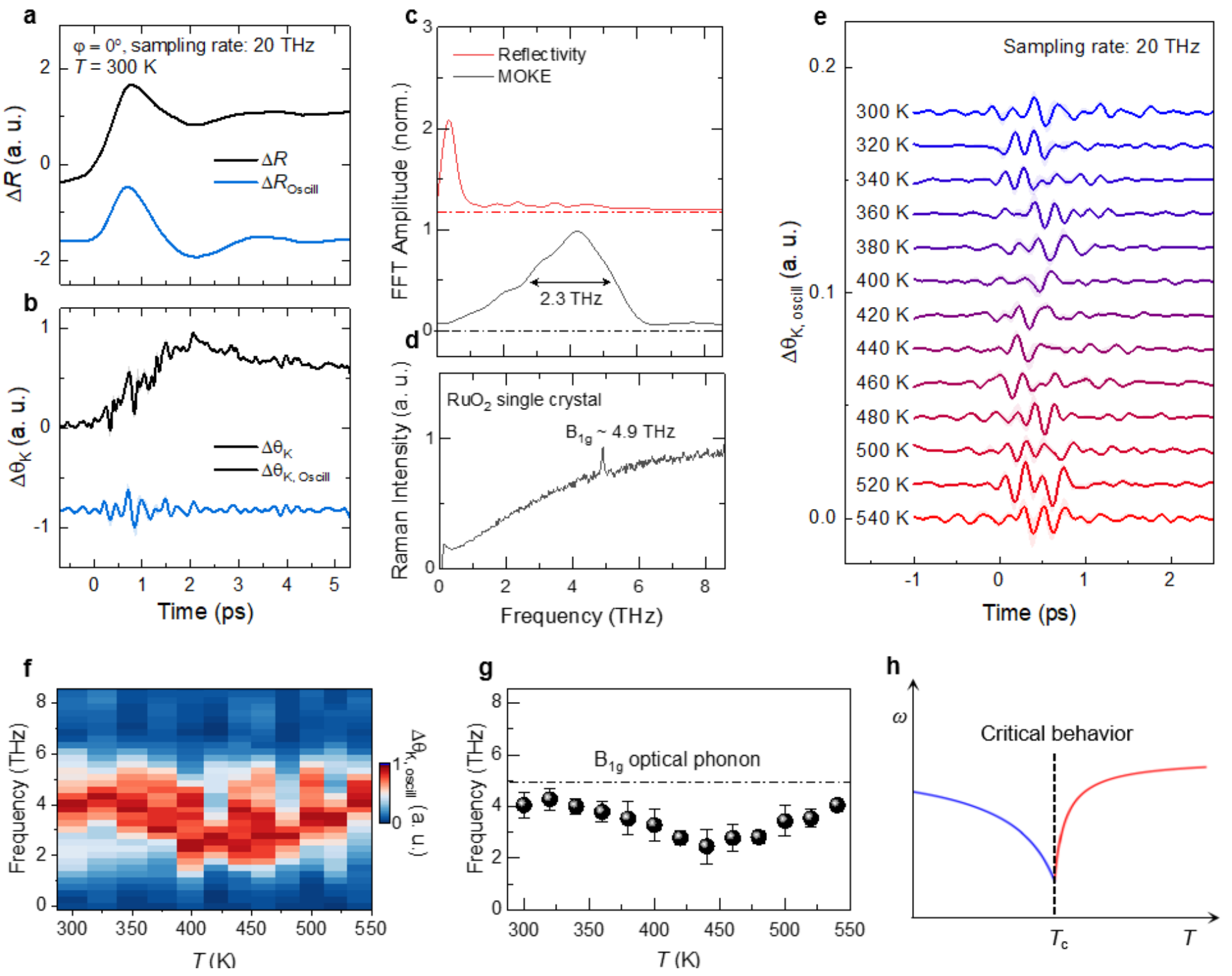

**Figure 4. Time-resolved magneto-optic Kerr effect measurements for 12 nm $RuO_2$ (110) film. (a, b)** Photo-induced $\Delta R$ (black line, top panel) and $\Delta\theta_K$ (black line, bottom panel), and their oscillation parts $\Delta R_{oscill}$ (blue line, top panel) and $\Delta\theta_{K,oscill}$ (blue line, bottom panel) obtained with the sampling rate of 20 THz where the frequency window is given up to 10 THz. **(c)** FFT spectra of $\Delta R_{oscill}$ (red) and $\Delta\theta_{K,oscill}$ (black). **(d)** Raman spectrum of $RuO_2$ single crystal exhibiting $B_{1g}$ phonon mode at 4.9 THz. **(e)** $T$-dependent $\Delta\theta_{K,oscill}$, and **(f)** the FFT spectra of them. **(g)** The center frequency of $B_{1g}$ mode as a function of $T$, exhibiting the notable frequency softening near $T$ = 450 K. **(h)** Schematic of renormalization of energy via the spin-phonon coupling showing the frequency softening at $T_c$.

# Strain-induced Dynamic Spin-phonon Coupling in Epitaxial $RuO_2$ Films

In Hyeok Choi[1,†], Seung Gyo Jeong[2,3†], Jae Hyuck Lee[4], San Kang[4], Sreejith Nair[2], Changyoung Kim[4], Dirk Wulferding[5], Bharat Jalan[*,2], Jong Seok Lee[*,1]

[1]*Department of Physics and Photon Science, Gwangju Institute of Science and Technology (GIST), Republic of Korea*

[2] *Department of Chemical Engineering and Materials Science, University of Minnesota−Twin Cities, Minneapolis, Minnesota 55455, United States*

[3]*Department of Physics, Hankuk University of Foreign Studies, Yongin 17035, Republic of Korea*

[4]*Department of Physics and Astronomy, Seoul National University, Seoul 08826, Republic of Korea*

[5]*Department of Physics and Astronomy, Sejong University, Seoul 05006, Republic of Korea*

**The table of contents**

## Section S1. Frequency Spectra of Coherent Oscillations in $RuO_2$ (110) Films

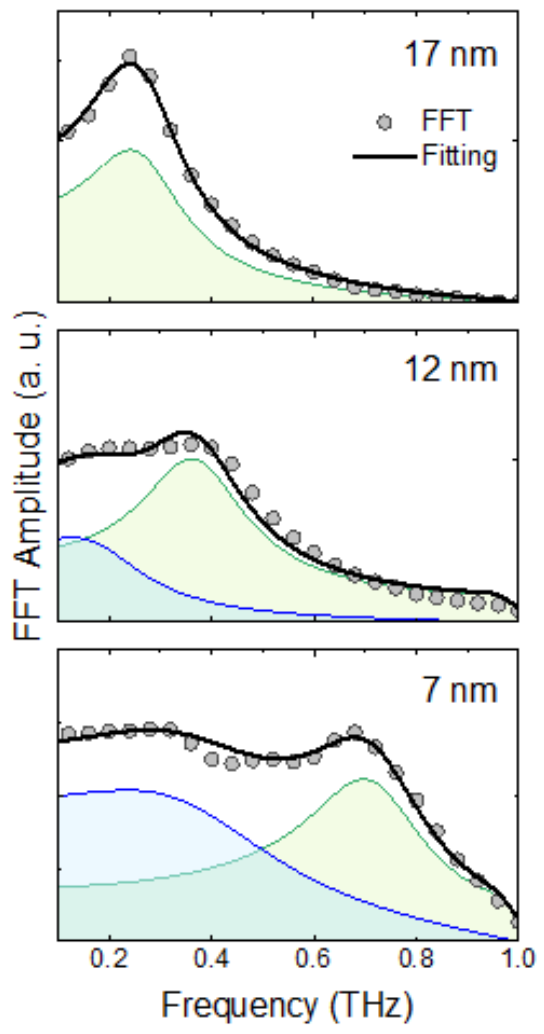

**Figure S1. Frequency-dependent amplitude spectra of coherent oscillations for 17, 12, and 7 nm $RuO_2$ (110) films.**

To determine the center frequency of coherent oscillation for 17, 12, and 7 nm $RuO_2$ (110) films, we used a Fast-Fourier-Transform (FFT) of transient reflectivity obtained in the time domain, as shown in Fig. S1. We observed the maximum peak at ~ 0.25 THz in the spectrum of the 17 nm $RuO_2$ film (top panel). As the thickness decreases to 17 nm and 7 nm, there is a redshift of the maximum peak (green curve). For 12 nm and 7 nm $RuO_2$ (110) films, the broad peaks (blue curve) appear at the lower frequency, which may originate from the remnant exponential component in the initial rise following the laser excitation. We fit these spectra with two Lorentzian functions indicated with green and blue curves, and took the peak of the green curve as the frequency of the optically-induced coherent oscillation.

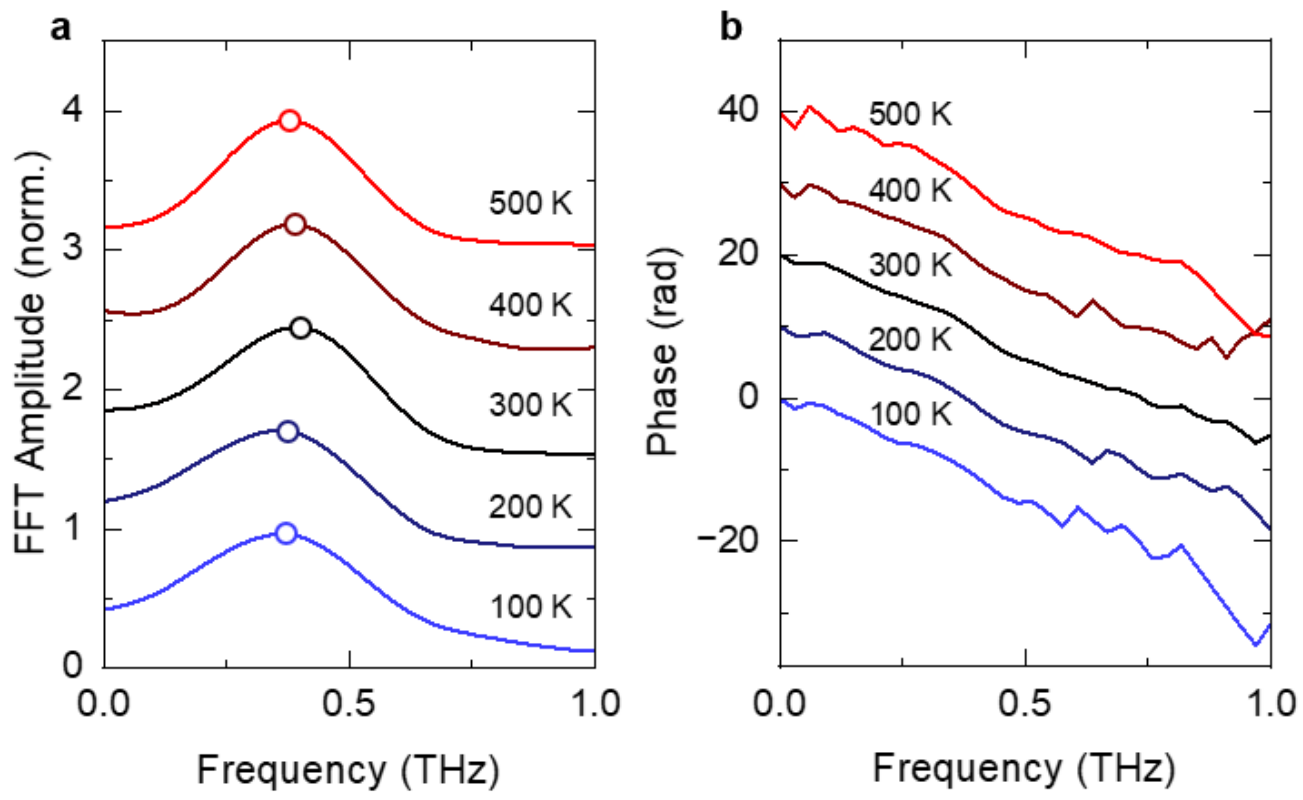

**Figure S2. FFT amplitude and phase spectra of THz coherent oscillations obtained at several temperatures.**

Figure S2 shows the FFT amplitude and phase spectra of THz coherent oscillations obtained at several temperatures. The FFT amplitude spectra exhibit a clear single peak with broad bandwidth (~ 0.3 THz), and their center frequencies are slightly modulated as temperature is changed (Fig. S2a). However, it is hard to obtain the information on time-dependent frequency softening or hardening in the FFT amplitude spectra. Furthermore, although the quadratic contributions in the FFT phase spectra (Fig. S2b) contain information on frequency softening or hardening, the linear term associated with the time shift is dominant, making an explicit analysis challenging.

## Section S2. GHz Coherent Oscillation Observed in $RuO_2$ (110) Films

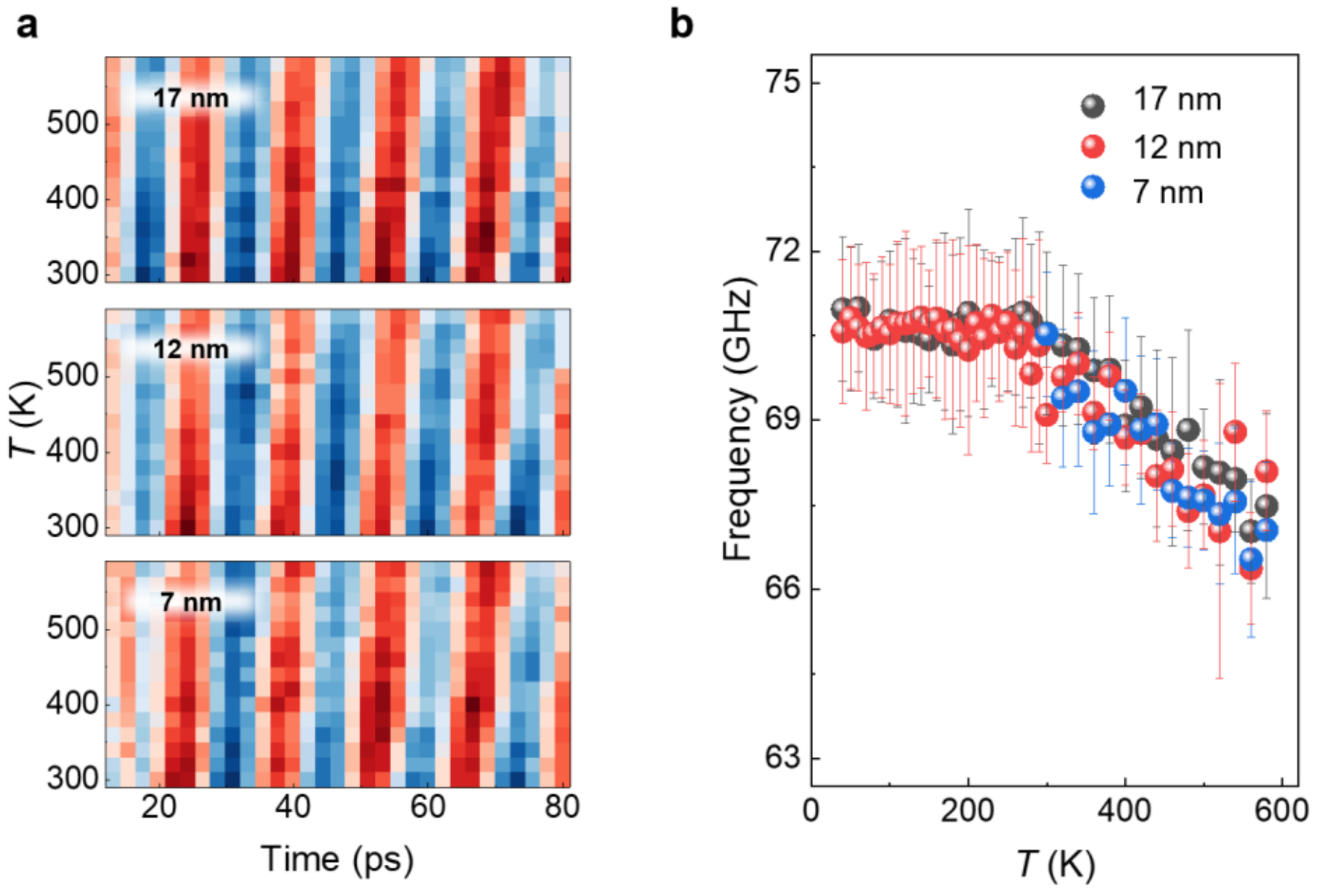

**Figure S3. Extracted GHz coherent oscillation of 17 nm, 12 nm, and 7 nm $RuO_2$ films as functions of time and temperature.**

From the pump-induced change in reflectivity $\Delta R$ for $RuO_2$ (110) films, we observed both THz and GHz oscillations. Figure S3a presents the extracted oscillation components $\Delta R_{oscill}$ as functions of time and temperature ($T$) for 17, 12, and 7 nm films. Their frequencies exhibit the same $T$-tendency regardless of thickness, as shown in Fig. S3b. This suggests that the GHz oscillation is induced by the interference of two beams reflected between the film surface and the strain wave propagating into the substrate. The frequency of interference is given as $2nv_s/\lambda$, where $\lambda$ is the wavelength of light, $n$ is the refractive index, and $v_s$ is the sound velocity of the substrate. At room temperature, $v_s$ along the [110] axis is $1\times10^4$ m/s, similar to the reported value of $TiO_2$ when the strain wave propagates along the [110] axis[1].

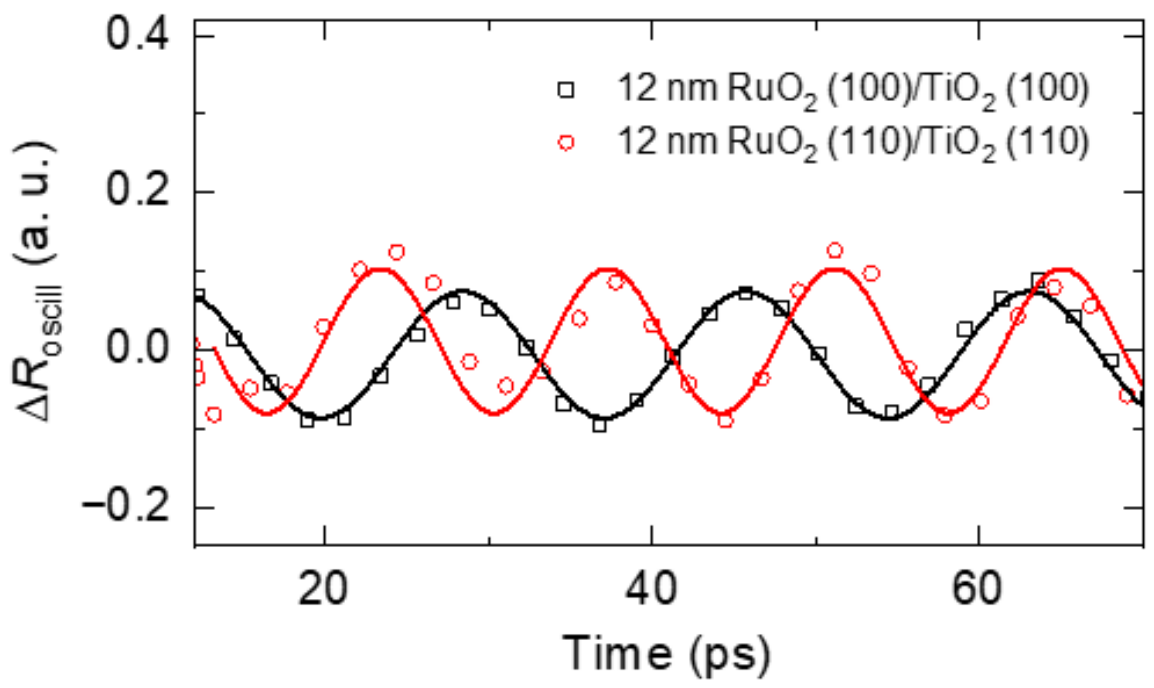

**Figure S4. Comparison GHz coherent oscillations observed in $RuO_2$ (100) and $RuO_2$ (110) thin films.**

Figure S4 compares the dynamics of coherent oscillations in $RuO_2$ (100) and $RuO_2$ (110) thin films obtained at $\varphi = 0°$ (probe || [001]). The film thickness is 11.5 nm and 12.5 nm, respectively, and the crystallographic orientation of the $TiO_2$ substrates corresponds to that of the respective $RuO_2$ films. Notably, the GHz coherent oscillations are given differently according to the crystallographic-orientation. To compare the frequency of the GHz oscillations, we fitted them with sinusoidal functions (solid lines), and the obtained frequencies are 57 GHz and 70 GHz, respectively. The corresponding longitudinal sound velocities are $7.6 \times 10^3$ m/s and $10 \times 10^4$ m/s for $TiO_2$ (100) and $TiO_2$ (110), respectively, consistent to theoretical expectation[1].

## Section S3. Pump and Probe Polarization-dependent THz Oscillation Amplitude of 12 nm $RuO_2$ (110) and (100) Films

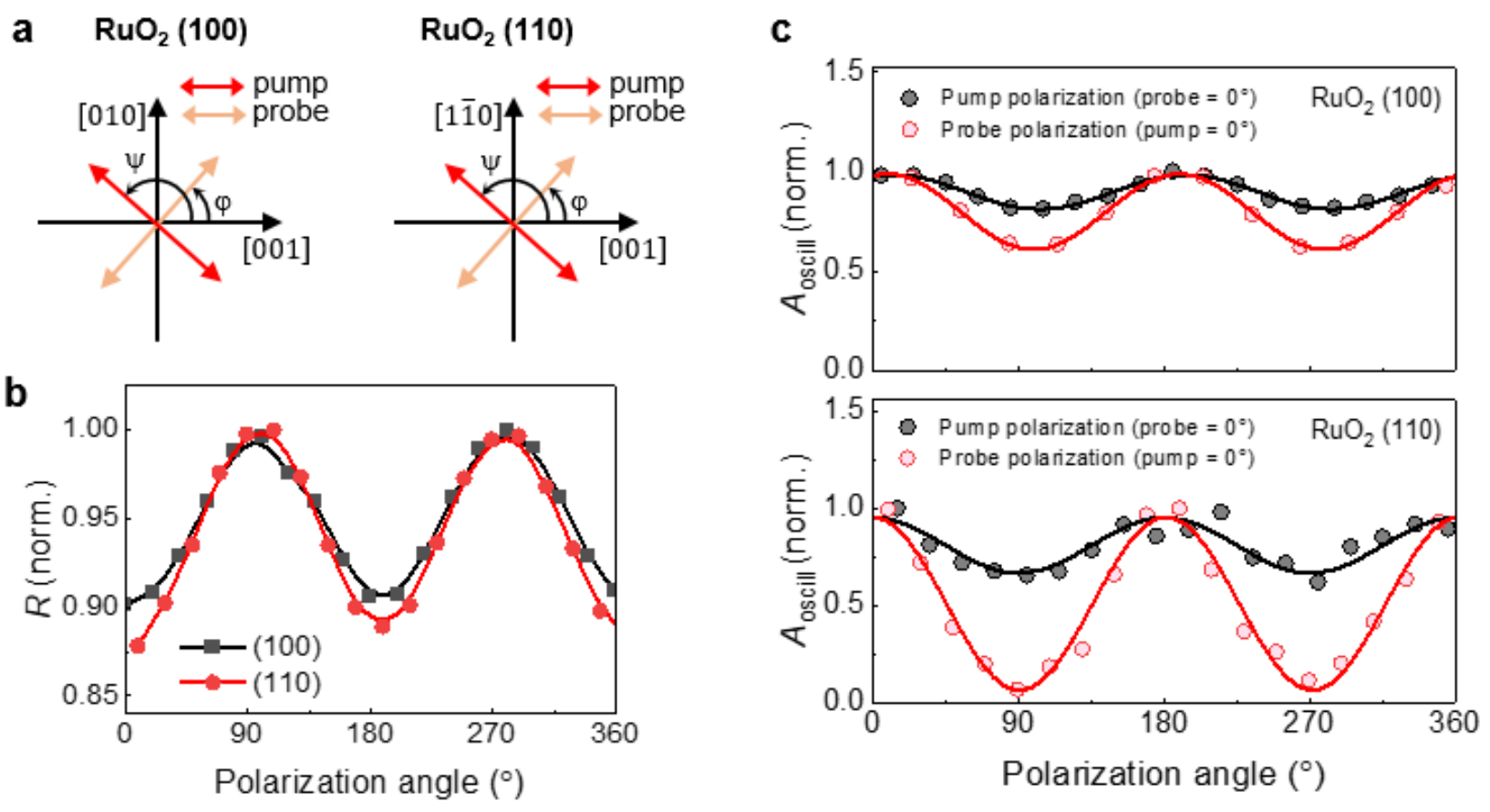

**Figure S5. Probe and pump polarization-dependent oscillation amplitude of the 12 nm $RuO_2$ thin film.**

To elucidate the polarization of coherent oscillation in 12 nm $RuO_2$ films grown on the two different crystallographic orientations (110) and (100) of $TiO_2$ substrates, we measured THz coherent oscillations by rotating the probe polarization angle (φ) and the pump polarization angle (ψ) (Fig. S5a). We note that both the probe and pump polarization angles are rotated from the [001] direction, i.e., φ = 0° (or ψ = 0°) corresponds to the [001] direction. Before obtaining the coherent oscillation, we first measured the change in the reflectivity $R$ with rotating φ for $RuO_2$ (110) and (100) films, as shown in Fig. S5b. We observed the clear difference in $R(\varphi)$ between [001] and [010] (or [1$\bar{1}$0]) directions for the $RuO_2$ (100) (or (110)) films due to their strongly anisotropic electronic structure[2]. Figure S5c shows φ- and ψ-dependent oscillation amplitude $A_{oscill}$ for the $RuO_2$ (100) and (110) films (top and bottom panels, respectively), normalized by $R$ obtained in Fig. S5b. When we measured φ-(or ψ-)dependent $A_{oscill}$, we fixed ψ (or φ) as 0°, and observed the two-fold in-plane anisotropic $A_{oscill}$ for both films. This suggests that the coherent oscillation corresponds to the transverse acoustic (TA) mode, where the atomic motion is perpendicular to the wave propagation [110] directions rather than the longitudinal acoustic (LA) mode. We observed the anisotropic nature also in the $RuO_2$ (100) film, even though it is weaker than that in the $RuO_2$ (110).

## Section S4. Elastodynamic Simulation in Strain-relaxed $RuO_2$ (110) Film

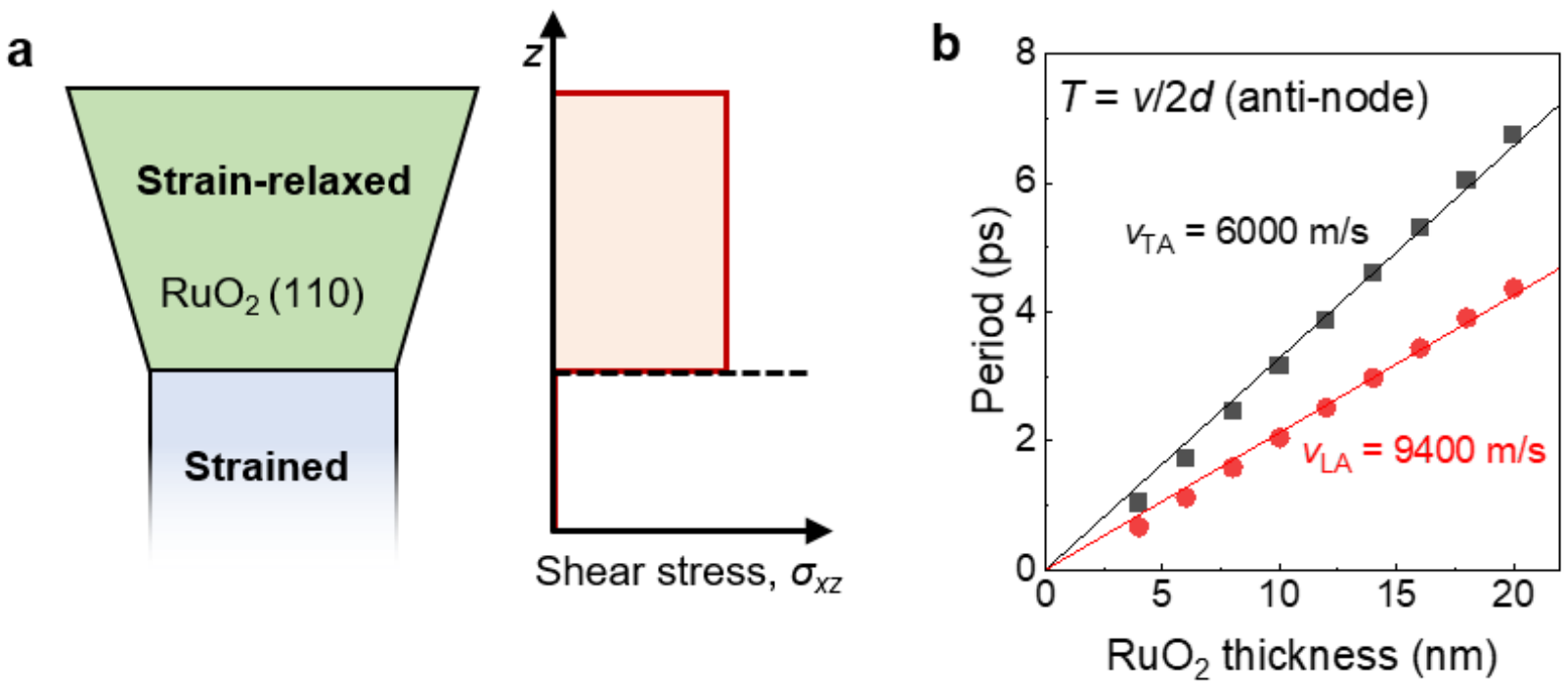

**Figure S6. Elastodynamic simulation for the thermal stress induced coherent standing elastic wave in the strain-relaxed $RuO_2$ thin film.**

In general, it is hard to generate transverse acoustic (TA) phonons in centrosymmetric media via optical excitation[3]. However, if there is a non-zero shear stress $\sigma_{xz}$, the small perturbation from the optical pump can lead to a change in $\sigma_{xz}$, inducing TA phonons in films. To examine the role of $\sigma_{xz}$, possibly related to the origin of TA phonons, we conducted the elastodynamic simulation for the $RuO_2$ film with non-zero $\sigma_{xz}$, as shown in Fig. S6. We assume that the $RuO_2$ film is separated into the strain-relaxed and strained heterostructures, which have non-zero and zero $\sigma_{xz}$, respectively. For the simulation, we adopted the staggered-grid finite method[4] by using a rectangular mesh whose size is 1 $nm^3$ and the number is $20 \times 20 \times d$, where the $RuO_2$ thickness is calculated by $d \times 1$ nm. As schematically shown in Fig. S6a, at the top and bottom end layers, we used the fixed boundary conditions, leading to the anti-node standing wave for both longitudinal acoustic (LA) and TA modes. Furthermore, at the bottom layer, we set non-zero $\sigma_{zz}$ from the discontinuity of the interface, which can induce the LA mode, while $\sigma_{xz}$ is given as a constant value for all layers. We use the periodic boundary condition for the side end layers. To solve the time-dependent elastodynamics equation, we adopted the $4^{th}$-order Runge-Kutta method which can provide much lower numerical errors compared to the Euler method. Figure S6b exhibits the simulated period of the TA and LA standing wave (black and red, respectively) as a function of $RuO_2$ film thickness.

In the upper panel of Fig. S7, sound velocity ($v_s$, green circle) increases as the film thickness increases, and it is eventually saturated to approximately 6000 m/s at 10 nm. The saturated $v_s$ matches well the calculated $v_s$ of TA mode indicated by the grid line. We note that the reduction of $v_s$ in thin films has been reported in several metallic films, consistent with our observation[5-7]. Assuming that the coherent oscillation is only defined in the strain-relaxed area in the films, we calculated the sound velocity $v_s$ using the standing wave equation $v_s = 2f \cdot (d - d_0)$, where $f$ is the frequency of the standing wave, $d$ is the thickness of the film, and $d_0$ is the averaged thickness of strained area (assuming to 4 nm). The estimated frequency of the standing wave $f$ is summarized as a red symbol in the upper panel. The bottom panel of Fig. S7 exhibits the thickness-dependent oscillation amplitude ratio $A_{\mathrm{oscill}}$ obtained by dividing the amplitude of exponential decay by that of oscillation. $A_{\mathrm{oscill}}$ is linearly reduced as the thickness decreases to 7 nm, and it becomes zero below 4 nm for THz oscillations. Interestingly, this characteristic thickness, where the sub-THz oscillations appear, is similar to the critical thickness of strain relaxation in $RuO_2/TiO_2$ (110) films (4-6 nm) observed in optical second harmonic generation and X-ray diffraction (XRD) results[8]. This implies the strain relaxation starts above 4 nm, so above 6 nm of $RuO_2$ (110) can coexist with strained and strain-relaxed states in the film[9]. On the other hand, $A_{\mathrm{oscill}}$ of GHz oscillation exhibits a non-zero value even below 6 nm, demonstrating the substrate origin.

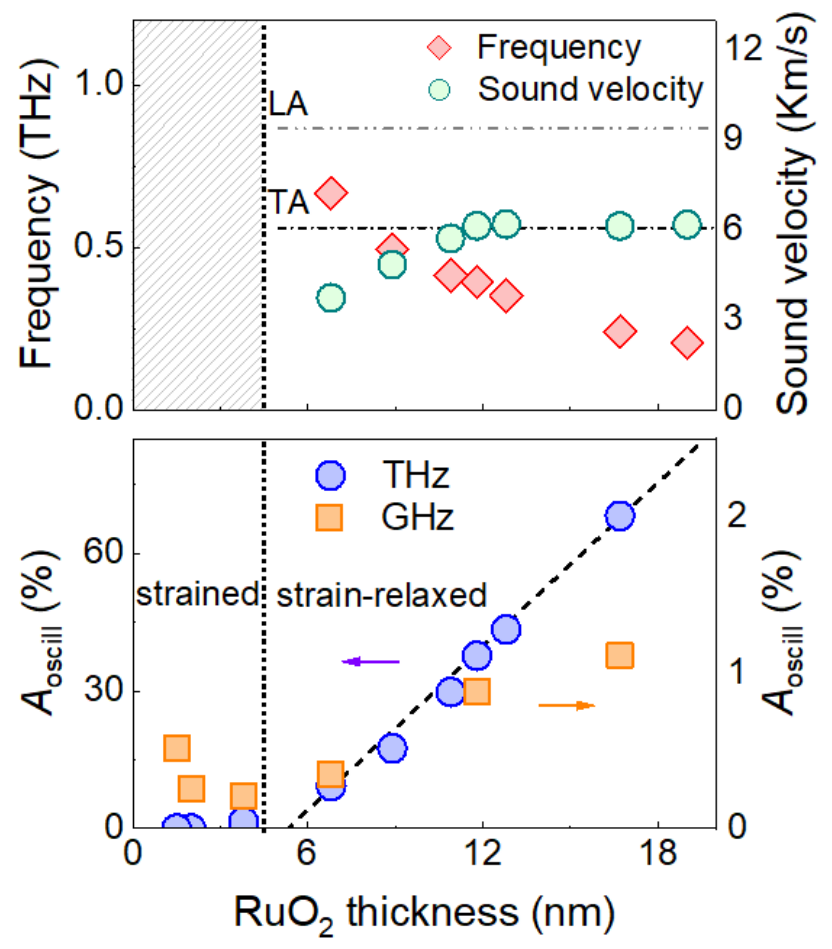

**Figure S7. Thickness-dependent sound velocity (top panel) and the amplitude of oscillation for $RuO_2$ (110) films.**

## Section S5. Oscillation Peak Position Fitting with the Quadratic Function

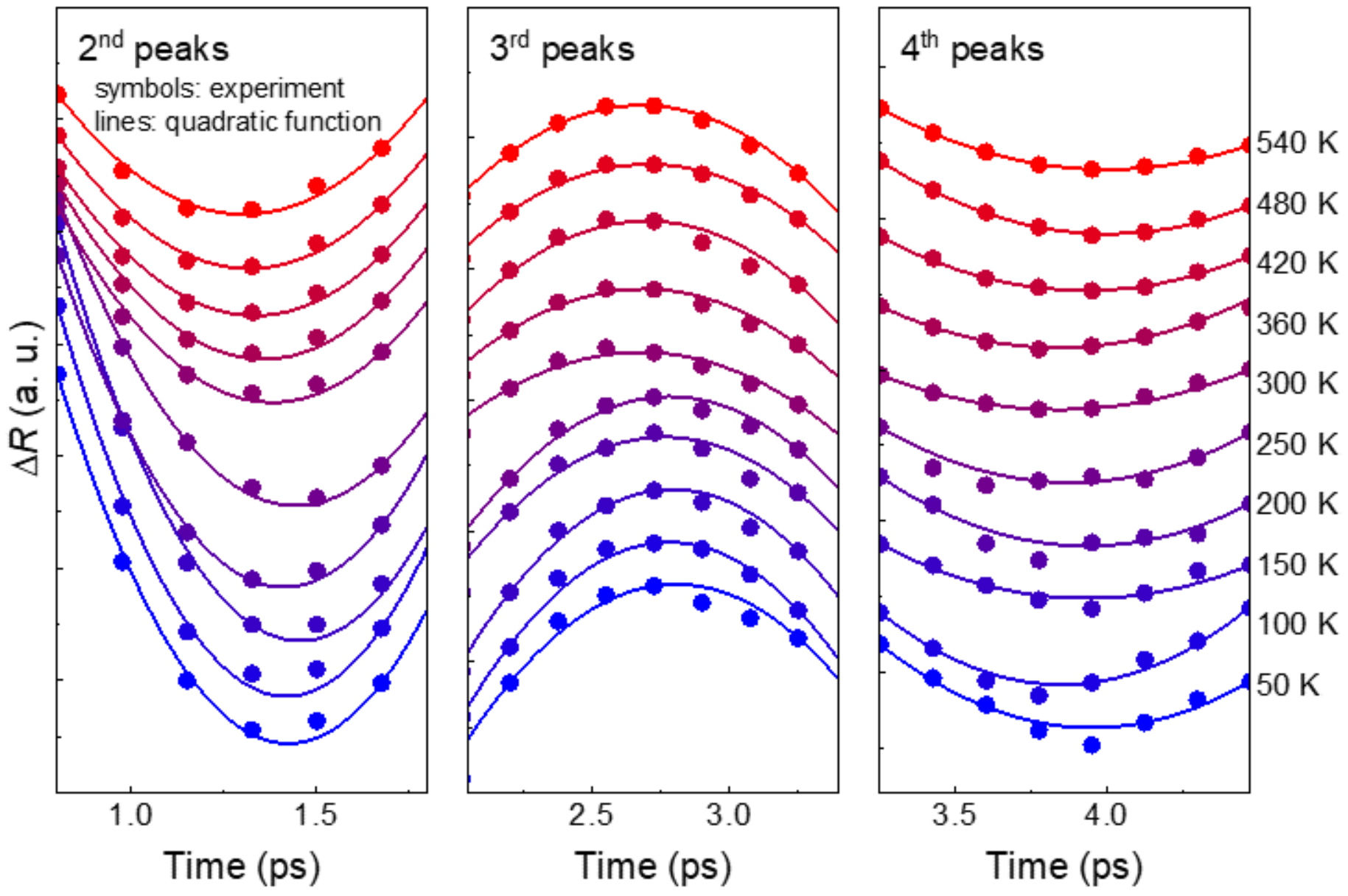

**Figure S8. Fitting the peak of coherent oscillations with the Lorentzian function for 12 nm $RuO_2$ (110) film.**

To obtain the time-varying oscillation frequency of a 12 nm $RuO_2$ (110) film, we fitted each maximum and minimum peak in the coherent oscillation with the quadratic function, as shown in Fig. S8. They exhibit the clear shift of peak positions as a function of temperature, demonstrating the possible correlation between TA modes and the magnetic state in the $RuO_2$ (110) films. Figure S9a describes an alternative method for determining extremal positions from the zero positons in the first derivative of signal. First, we obtained the first derivative of the coherent oscillation d$\Delta R$/d$t$, as displayed in the bottom panel for the representative result obtained at $T$ = 580 K (top). By fitting d$\Delta R$/d$t$ with the linear function (red), we can reasonably obtain the zero positon for the maxima and minima position P1, P2, and P3 (dashed lines).

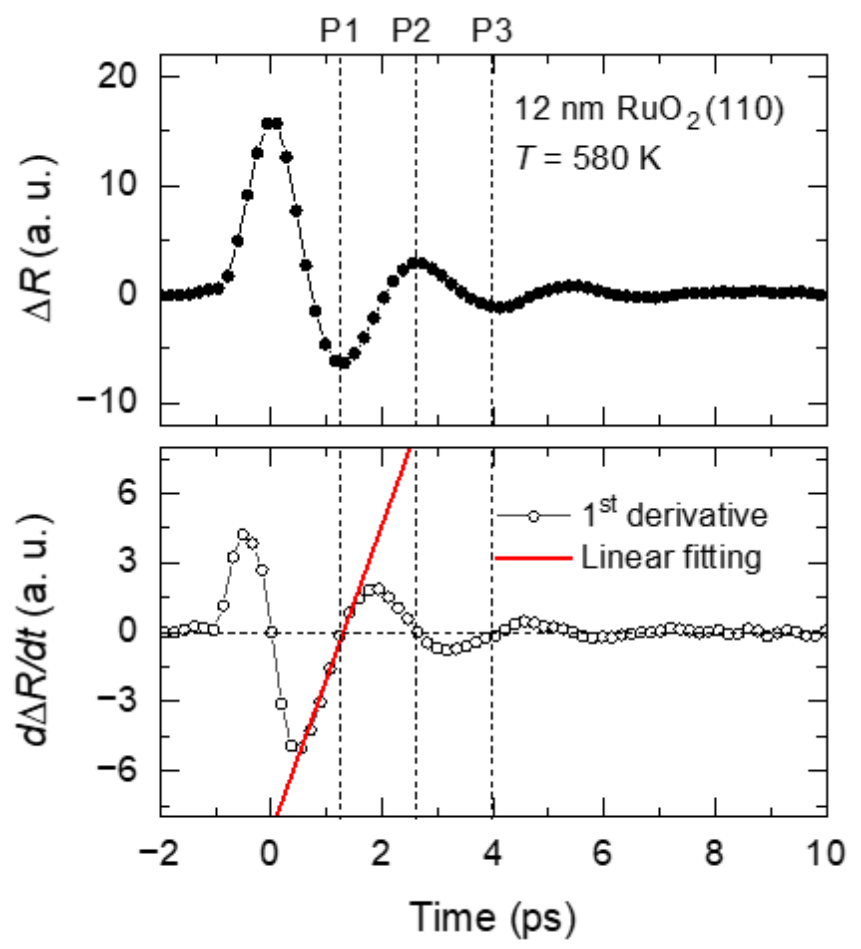

**Figure S9. Determination of the maxima and minima positions from the derivative of the oscillation signals.**

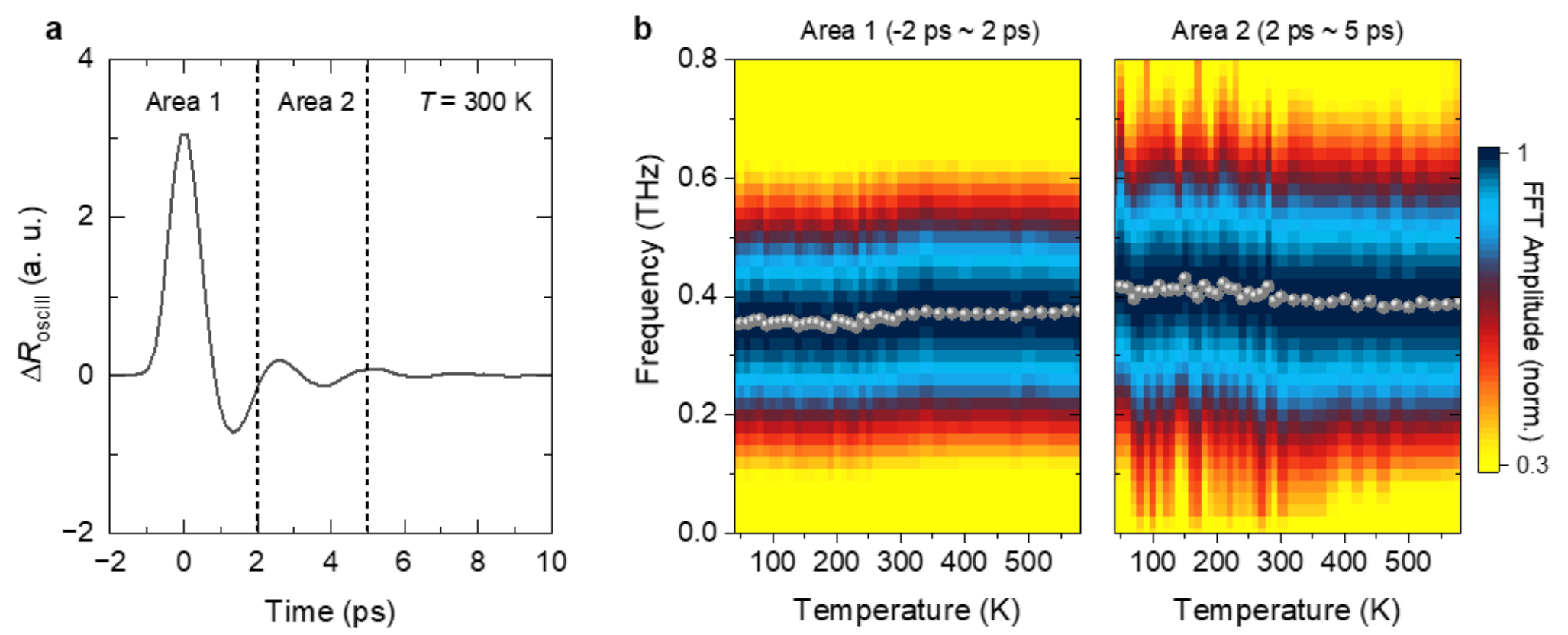

**Figure S10. FFT amplitude spectra obtained at area 1 and area 2.**

To verify the validity of time-domain-based methods, we additinoally determined the peak positions from the FFT spectra obtained from two distinct regions: area 1 (from –2 to 2 ps) and area 2 (from 2 to 5 ps) (Fig. S10a). As shown in Fig. S10b, the peak position in FFT spectra for both area 1 and 2 exhibits a temperature dependence, while their temperature trend is opposite upon heating: frequency hardening for area 1 and softening for area 2. Figure S11 summarizes the obtained peak positions from FFT (a), and from time-domain methods 1 (b) and 2 (c), empolying the original oscillation singals and the first derivative of the singals, respectively. All methods consistently provide similar temperature-dependent peak positions, confirming the strong temperature-dependent changes in THz coherent oscillations in $RuO_2$ (110) films.

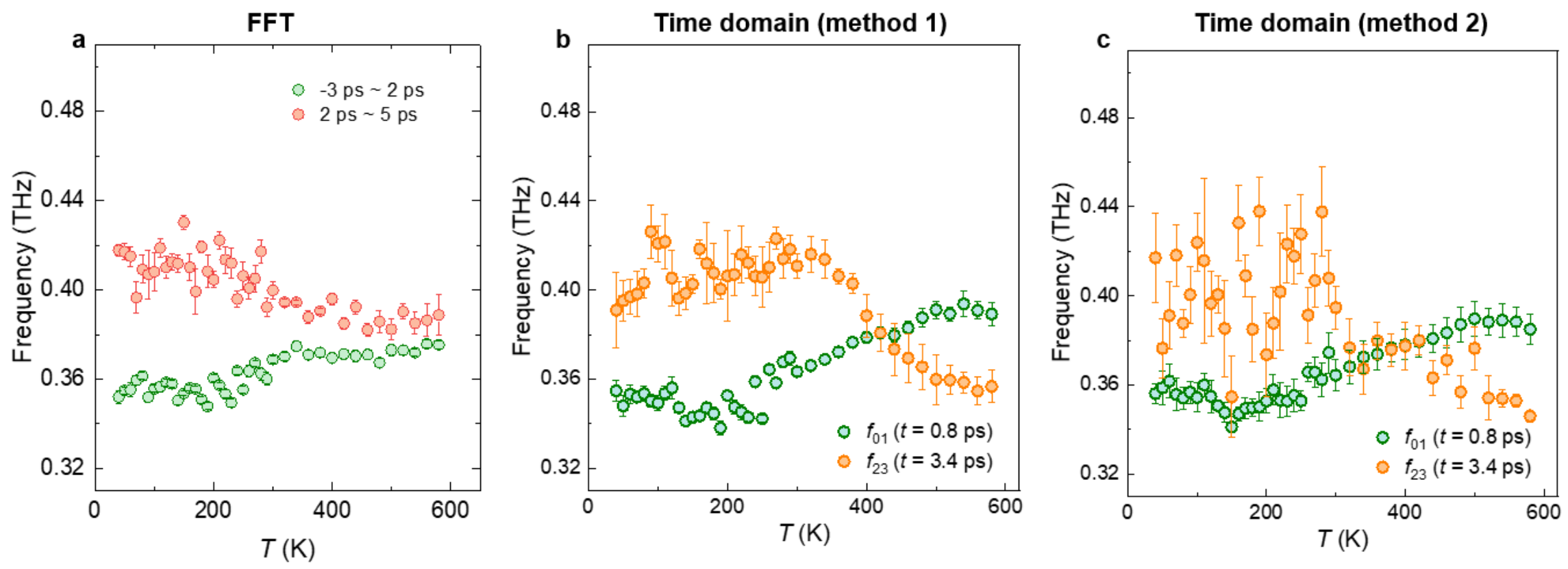

**Figure S11. Comparison temperature-dependent quasi-static frequencies obtained in a) frequency-domain and b) time-domain.**

## Section S6. Two-Temperature Model Analysis

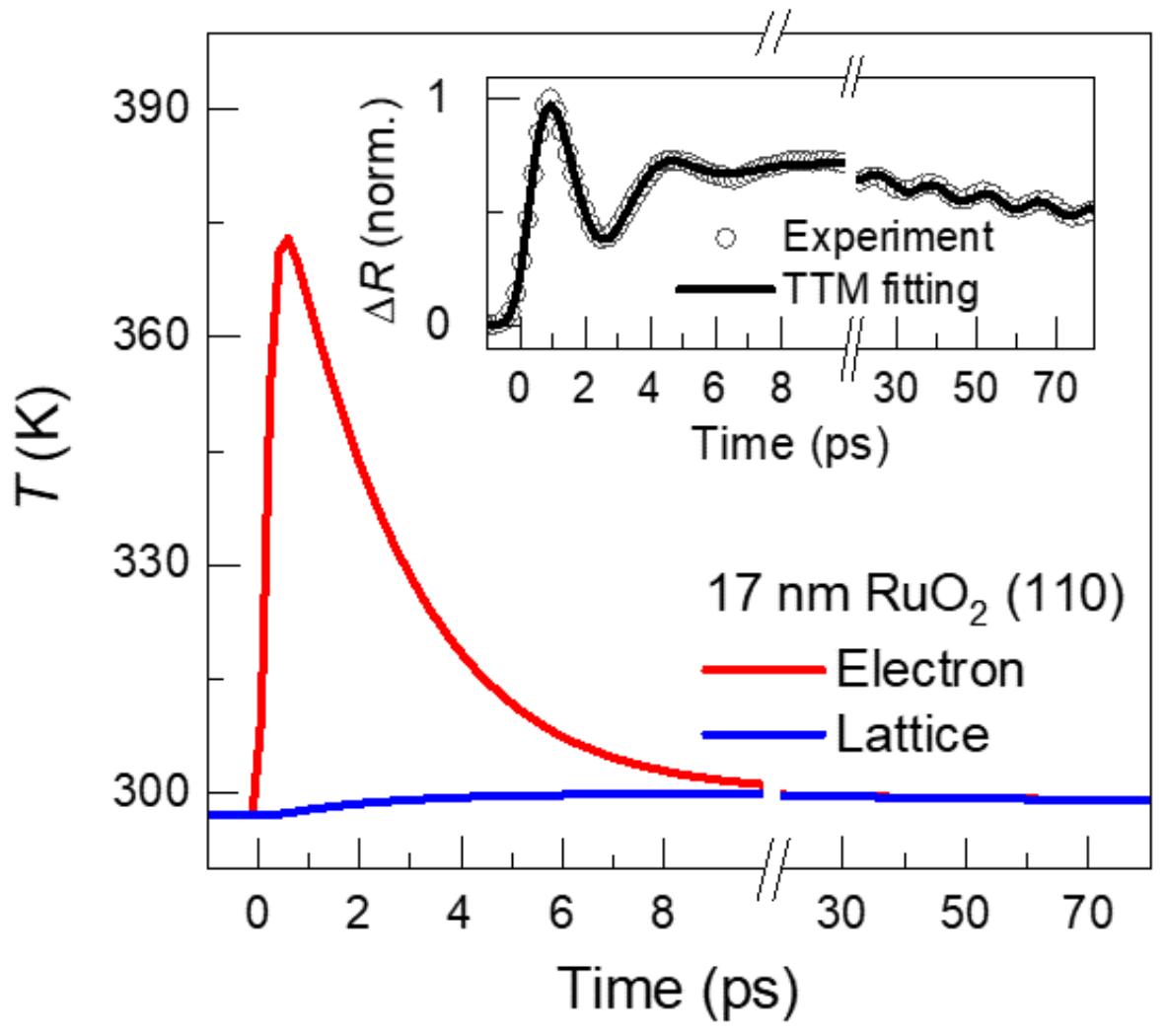

**Figure S12. Two-temperature model analysis for 17 nm $RuO_2$ (110) film to simulate the transient electron temperature (red) and lattice temperature (blue).**

To simulate the evolution of temperature changes in electron and lattice systems, we conducted two temperature model (TTM) analyses[10]. For the simulation, we employed the one-dimensional finite difference method with a 1 nm grid size. We set the optical penetration depth, heat capacity, and thermal conductivity of $RuO_2$ films as 26 nm, 2.88 J/cm$^3$ K, and 11 W/m K, respectively[11,12]. At the film surface and the end of the substrate, the Neumann boundary condition and semi-infinite boundary condition are applied, respectively. By fitting $\Delta R$ of 17 nm $RuO_2$ film with TTM combined with sine oscillations, spatial distributions of electron temperature ($T_e$) and lattice temperature ($T_l$) are obtained as a function of time using the Euler method with 10 fs time step, while the electron-phonon coupling constant $G$ is given by ~ $5 \times 10^{16}$ W/m$^3$ K (Fig. S12).

In the main text, we estimated the frequency shift $f_{01}$–$f_{23}$ at room temperature based on the TTM simulation results and the experimental results shown in Fig. 3c. We assume that the TA phonon frequency $f_{23}(T_e)$ is determined by the instantaneous electronic temperature $T_e$. Since the highest $T_e$ is approximately 380 K obtained from the TTM simulation, the photo-induced phonon frequency shift $f_{01}$–$f_{23}$ at room temperature can be estimated as $f_{23}$(380 K)–$f_{23}$(297 K) ≈ 0.39 – 0.42 THz = –0.03 THz, which is in good agreement with experimentally obtained value of $f_{01}$–$f_{23}$ at room temperature (≈ –0.04 THz).

## Section S7. Lattice Heating Effect on Time-Varying Mode Softening and Hardening

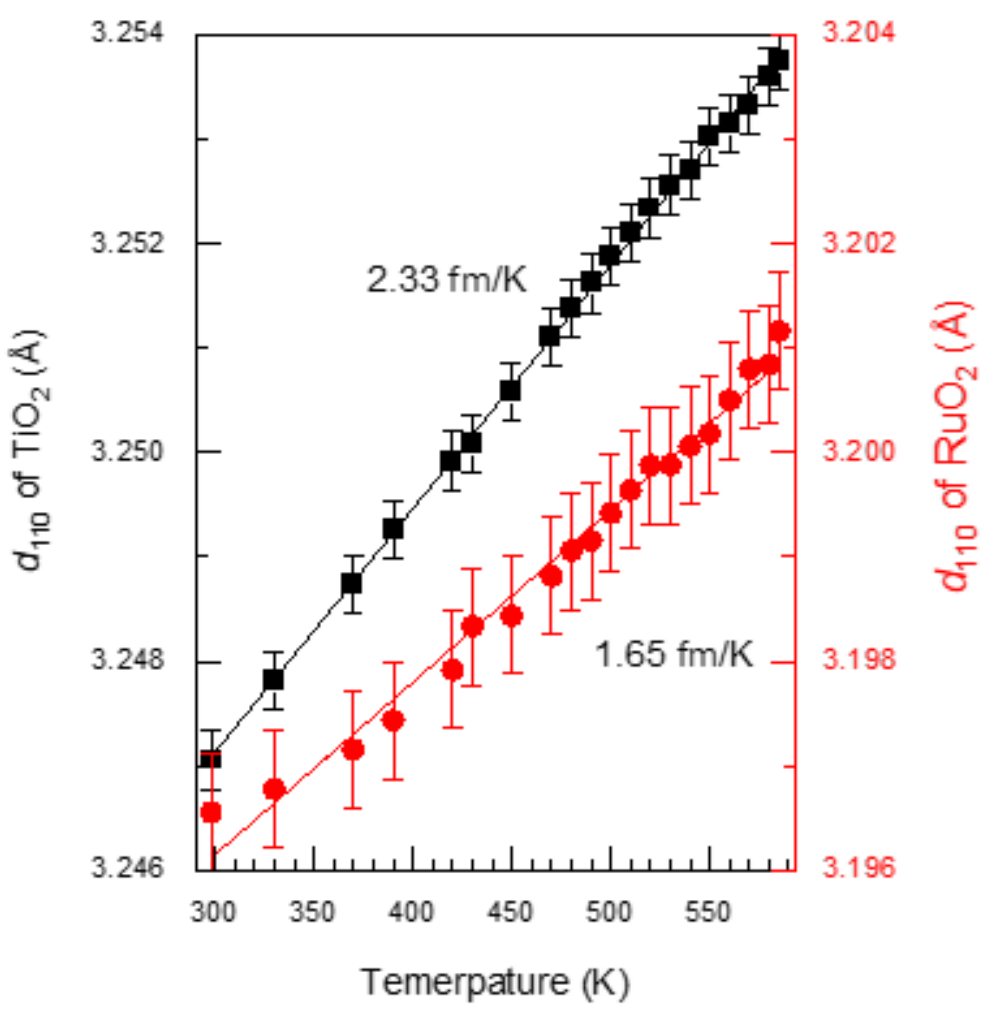

**Figure S13. Temperature-dependent change in the lattice constant along the [110] direction $d_{110}$ for $TiO_2$ (black) and $RuO_2$ (red).**

Although the lattice heating effect upon photoexcitation can affect the frequency of the standing wave, we here show that such effect is minimal in our measurement results. First, we discuss the dynamic change in the oscillation frequency induced by the increase in lattice temperature. To our knowledge, temperature-dependent stiffness tensor components of $RuO_2$ have not yet been reported. Therefore, we estimate their effects using our experimental results for $TiO_2$ (Section S2) which shares a similar rutile crystal structure and comparable Debye temperature (~600 K for $TiO_2$ and ~640 K for $RuO_2$). Notably, the sound velocity of $TiO_2$ decreases linearly upon heating from 250 K, with a slope of –2 m/s·K. Based on our two-temperature model simulation (Section S6), the elevated lattice temperature ($\Delta T_L$) is approximately 2 K, resulting in a decrease of about 4 m/s in the sound velocity. This corresponds to a −0.04% change in the sound velocity at $T = 400$ K. Since the frequency of coherent phonon oscillations $f$ in $RuO_2$ thin films is proportional to the sound velocity $v_s$ and the thickness $d$ ($f \sim v_s d$), the modulation of the sound velocity due to the increase in lattice temperature has a negligible effect on the significant frequency softening or hardening observed in our measurements (~ 10%).

Finally, we discuss the thermal expansion effect due to increase in lattice temperature. Figure S13 shows the temperature-dependent lattice constant along the [110] direction ($d_{110}$) in the $TiO_2$ substrate (black) and 17 nm $RuO_2$ thin film on $TiO_2$ (110) (red) obtained by using X-ray diffraction. The sample heating stage of the SmartLab XE with a spherical X-ray window (DHS-1100) under atmospheric conditions is used for these measurements. The sample was heated to 600 K and then cooled down to 300 K, while the positions of the (110) Bragg reflection peaks for each $RuO_2$ film and $TiO_2$ substrate were tracked as a function of temperature. The $d_{110}$ for both $TiO_2$ and $RuO_2$ is linearly proportional to temperature, indicating the positive thermal expansion. By fitting the data with linear functions, we obtained a slope of 1.65 fm/K for $RuO_2$, and this corresponds to a total increase in the $RuO_2$ layer thickness of approximately 0.001% under photoexcitation. Consequently, the purely structural/thermal effects cannot account for our observation of a 10% dynamic modulation in the coherent phonon frequency, indicating that the contributions from the electronic or spin system should be considered.

## Section S8. Pump-Power-Dependent THz Coherent Oscillation for 17 nm $RuO_2$ (110) Film

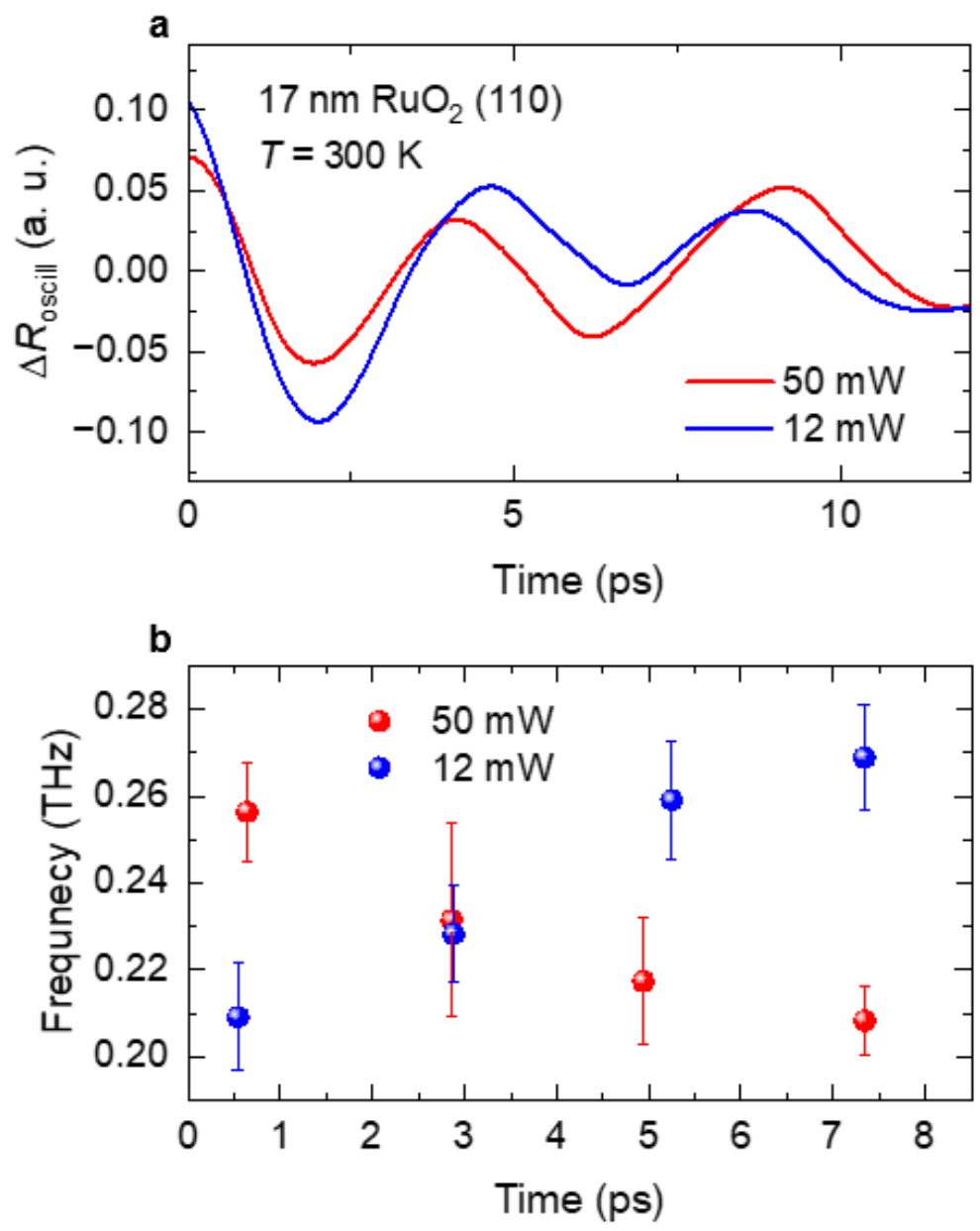

**Figure S14. Pump-power-dependent THz coherent oscillation meausred in 17 nm $RuO_2$ film.**

Under low fluence laser excitation, where the elevated electronic temperature remains below the transition temperature, the quasi-frequency of the TA phonon at room temperature decreases immediately after excitation and then recovers during cooling. This results in the transient frequency hardening described in the main text. In contrast, when the fluence is sufficiently high to raise the electronic temperature above the transition temperature, the frequency-hardening effect associated with spin-phonon coupling no longer dominates the dynamics. This is because, above the transition temperature, the quasi-steady-state TA phonon frequency exhibits only a weak temperature dependence. (Fig. 3c in the main text). In this case, the transient frequency softening will be observed, similar to the data obtained at high temperature (i.e. 580 K as shown in Fig. 3b), where the electron-phonon coupling itself plays a dominant role in determining the dynamic modulation of the phonon frequency.

To comfimr this expectation, we measured THz coherent oscillations at room temperature under two excitation powers, 12 mW (blue) and 50 mW (red), as shown in Fig. S14a. Here, we employed a 1 kHz repetition laser system to avoid the sample damage from the steady-state heating effect. The pump power of 12 mW corresponds to a laser fluence similar to that employed in the main text. Based on our two-temperature model simulations, the electronic temperature increase is estimated to be approximately 70 K and 280 K from the initial room temperature for the pumping power 12 mW and 50 mW, respectively. Figure S14b presents the quasi-frequency of the TA phonon as a function of time estimated by following the same process introduced in the main text; the TA phonon oscillation frequency exhibits hardening under 10 mW pumping, but it shows softening under 50 mW pumping. It should be noted that these behaviors are consistent with aforementioned expectations considering the spin-phonon and electron-phonon coupling, respectively, as a major contribution in each case of the low and high pumping power. Consequently, this clear power dependence in the dynamics of TA phonon oscillation strongly supports the presence of spin-phonon coupling in the experimental results obtained with the relatively low pumping power.

## Section S9. Field-dependent Time-resolved Magneto-optic Kerr Effect for 12 nm $RuO_2$ (110) Film

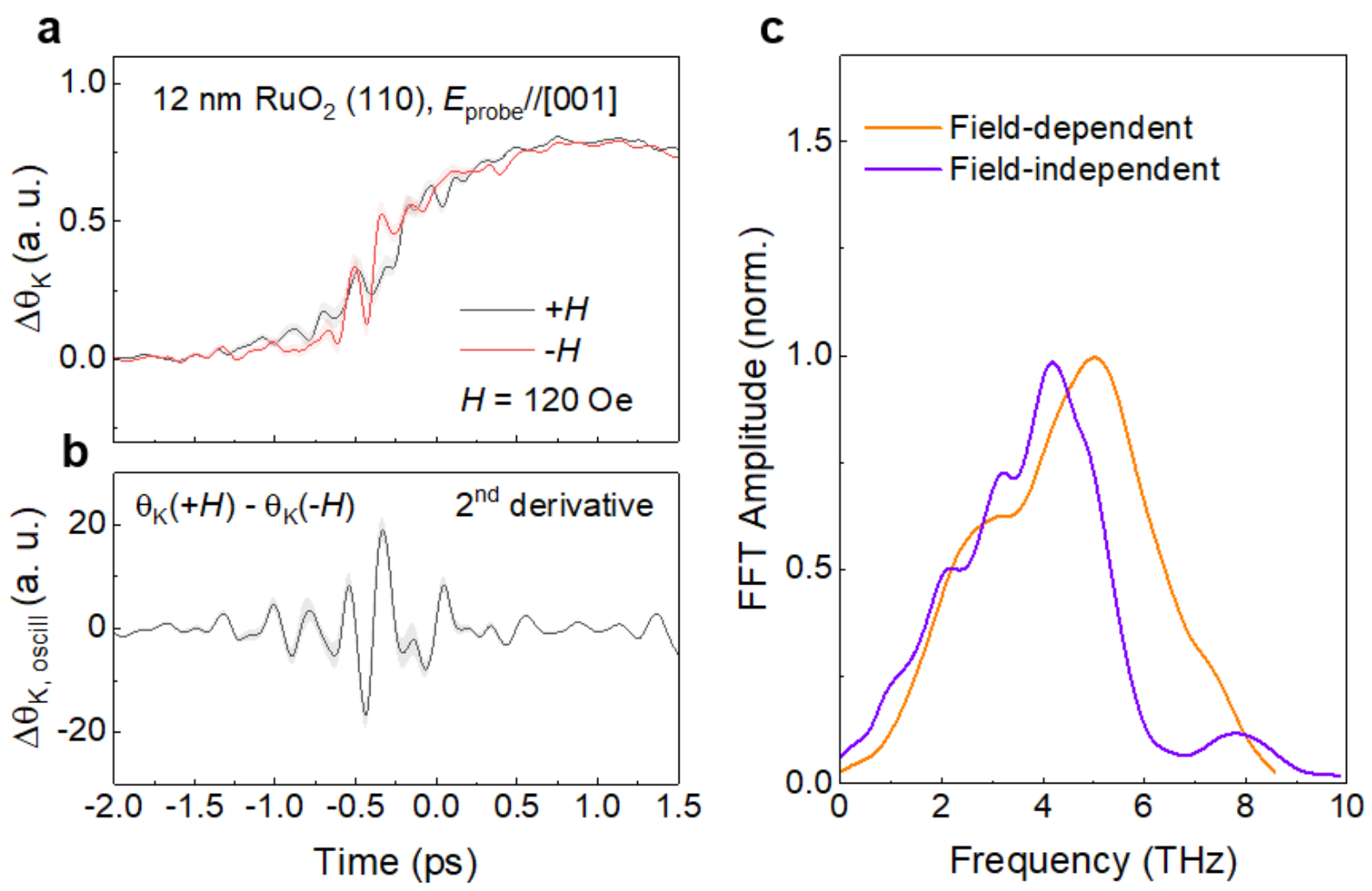

**Figure S15. Magnetic-field-dependent dynamics of magnetization for 12 nm $RuO_2$ (110) film.**

To examine the magnetic-field-dependent dynamics of magnetization, we measured $\Delta\theta_K$ with probe polarization along the [001] axis under the magnetic field $H$ of 120 Oe applied along the out-of-plane direction, as shown in Fig. S15a. We found that an exponential component is not changed according to the direction of $H$, demonstrating that the exponential part is not directly attributed to the field-induced magnetization. On the other hand, the THz oscillation components between $t$ = –1.0 and 0.5 ps are dependent on the direction of $H$, implying the THz oscillation is related to the dynamics of magnetization. Figure S15b exhibits the second-derivative of a field-dependent THz oscillation obtained by subtracting $\Delta\theta_K(+H)$ and $\Delta\theta_K(-H)$, corresponding to the pure magnetic component. We observed the meaningful oscillation, which is larger than measurement errors, depicted with the gray shaded line. The FFT spectrum (Fig. S15c) exhibits a single peak with a center frequency of 5 THz, corresponding to the $B_{1g}$ phonon mode. This demonstrates that the $B_{1g}$ phonon mode emerges in the dynamics of magnetization.

## Section S10. Phonon symmetry assignment via Raman spectroscopy on $RuO_2$ single crystals

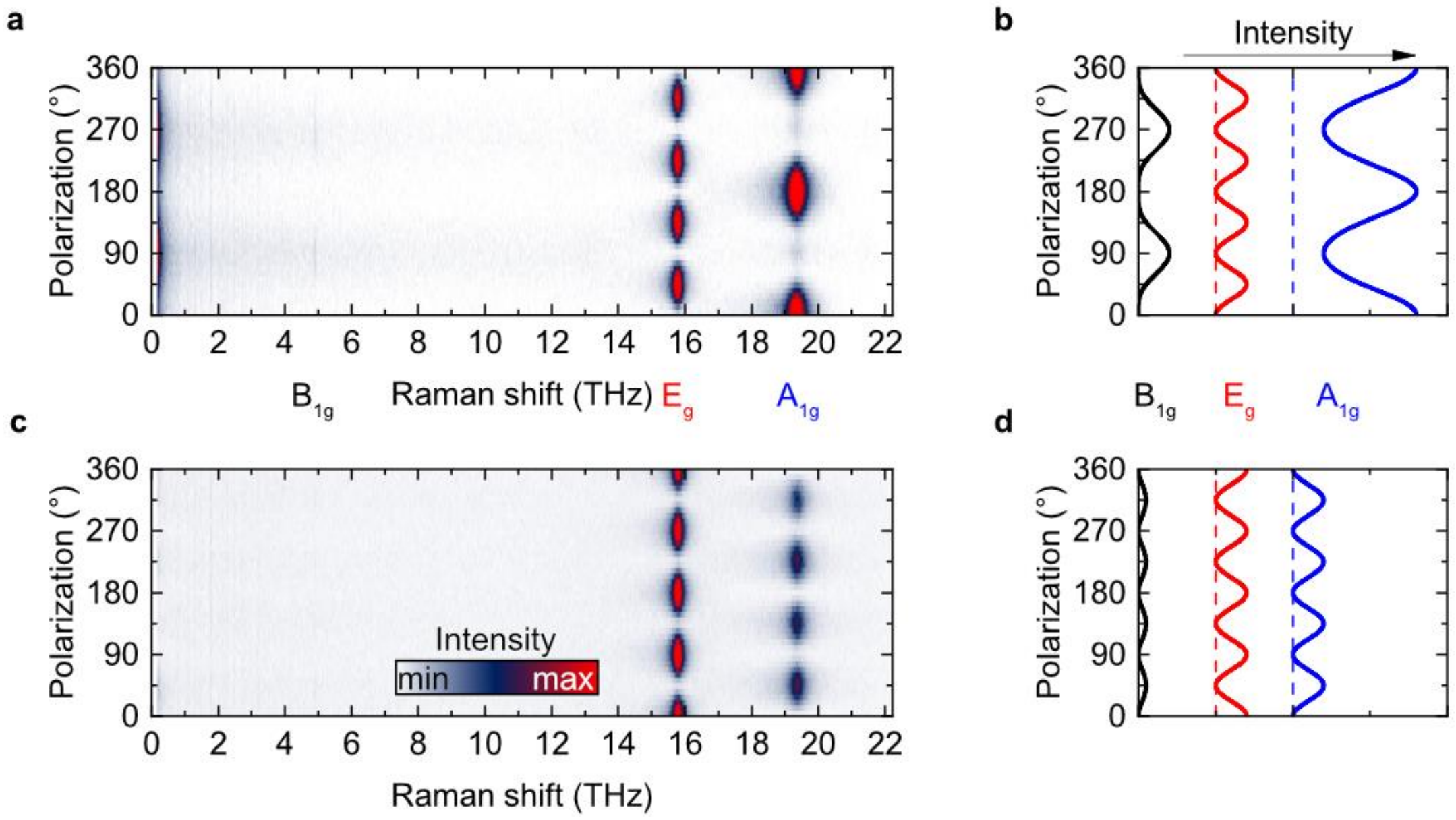

**Figure S16. Raman spectra taken on the (100) surface of a $RuO_2$ single crystal. (a) Polarization-resolved data measured in parallel configuration ($e_{in}$ // $e_{out}$). (b) Polarization-dependent intensity profiles calculated for $e_{in}$ // $e_{out}$. (c) Polarization-resolved data measured in crossed configuration ($e_{in} \perp e_{out}$). (d) Polarization-dependent intensity profiles calculated for $e_{in} \perp e_{out}$.**

To assign the symmetry of individual optical phonons in $RuO_2$, and to confirm the $B_{1g}$ symmetry of the mode at 4.9 THz, we conducted polarization-resolved Raman spectroscopy measurements on the (100) surface of a $RuO_2$ single crystal. These experiments were carried out at room temperature using a home-built microscope stage and a λ = 561 nm laser (*Oxxius*). The laser light was focused onto the sample with a spot diameter of 2 μm and a laser power of less than 0.3 mW. Light polarization was controlled via superachromatic λ/2 waveplates (*Thorlabs*), and the inelastically scattered light was dispersed through a Princeton Instruments TriVista spectrometer onto a PyLoN CCD detector. The measured intensity in a Raman scattering experiment is given by $I = |e_{in} \cdot R \cdot e_{out}|^2$, where $e_{in}$ and $e_{out}$ are the light polarization of the incident and outgoing photons, respectively, and $R$ is the Raman tensor of a given Raman-active mode. Hence, by controlling both $e_{in}$ and $e_{out}$ we can directly measure the Raman tensors and thereby access the symmetries of individual phonons.

In $RuO_2$, which crystallizes in the tetragonal space group $P4_2/mnm$ (#136), the Ru atoms are situated at the 2a Wyckoff position, while the O atoms reside at 4f. This configuration yields the following four Raman-active phonon modes: 1 $A_{1g}$ + 1 $B_{1g}$ + 1 $B_{2g}$ + 1 $E_g$. Their Raman tensors are given as

$$A_{1g}=\begin{pmatrix}a&0&0\\0&a&0\\0&0&b\end{pmatrix}, B_{1g}=\begin{pmatrix}c&0&0\\0&-c&0\\0&0&0\end{pmatrix}, B_{2g}=\begin{pmatrix}0&d&0\\d&0&0\\0&0&0\end{pmatrix}, E_g=\begin{pmatrix}0&0&0\\0&0&e\\0&e&0\end{pmatrix}.$$

We start by rotating the polarization of the light by 360° within the (100) plane of the single crystal, while keeping the relation $e_{\mathrm{in}}$ // $e_{\mathrm{out}}$. The results are shown as a color contour plot in Fig. S16a. We identify 2 intense modes at 15.8 THz and 19.4 THz, while a third, faint mode is seen at 4.9 THz. Based on the symmetry of the Raman tensors, we calculate the expected polarization dependence for experiments within the (100) plane. The resultant intensity plots are shown in Fig. S16b. Note that modes of $B_{2g}$ symmetry cannot be accessed in this scattering geometry, which leaves a total number of three Raman-active phonons. Comparing our experimental results to the calculated intensity plots, we find that the modes at 4.9 THz, 15.8 THz, and 19.4 THz are fully consistent with the $B_{1g}$, $E_g$, and $A_{1g}$ symmetry, respectively. As a consistency check, we repeated the experiment with $e_{\mathrm{in}} \perp e_{\mathrm{out}}$. The experimental data are shown as a color contour plot in Fig. S16c, while the calculated intensity profiles are plotted in Fig. S16d. Although the overall intensity of the $B_{1g}$ phonon at 4.9 THz is now too weak to trace, this second, independent dataset confirms our assignment for the 15.8 THz mode as $E_g$, and for the 19.4 THz mode as $A_{1g}$.

On the other hand, when the Raman tensors are transformed for the (110) plane of the single cystal, corresponding Raman tensors for $A_{1g}$, $B_{1g}$, $B_{2g}$ and $E_g$ modes are given as

$$A_{1g} = \begin{pmatrix} b' & 0 & 0 \\ 0 & a' & 0 \\ 0 & 0 & b' \end{pmatrix}, B_{1g} = \begin{pmatrix} 0 & 0 & c' \\ 0 & 0 & 0 \\ c' & 0 & 0 \end{pmatrix}, B_{2g} = \begin{pmatrix} d' & 0 & 0 \\ 0 & 0 & 0 \\ 0 & 0 & e' \end{pmatrix}, E_g = \begin{pmatrix} 0 & f' & 0 \\ f' & 0 & f' \\ 0 & f' & 0 \end{pmatrix}.$$

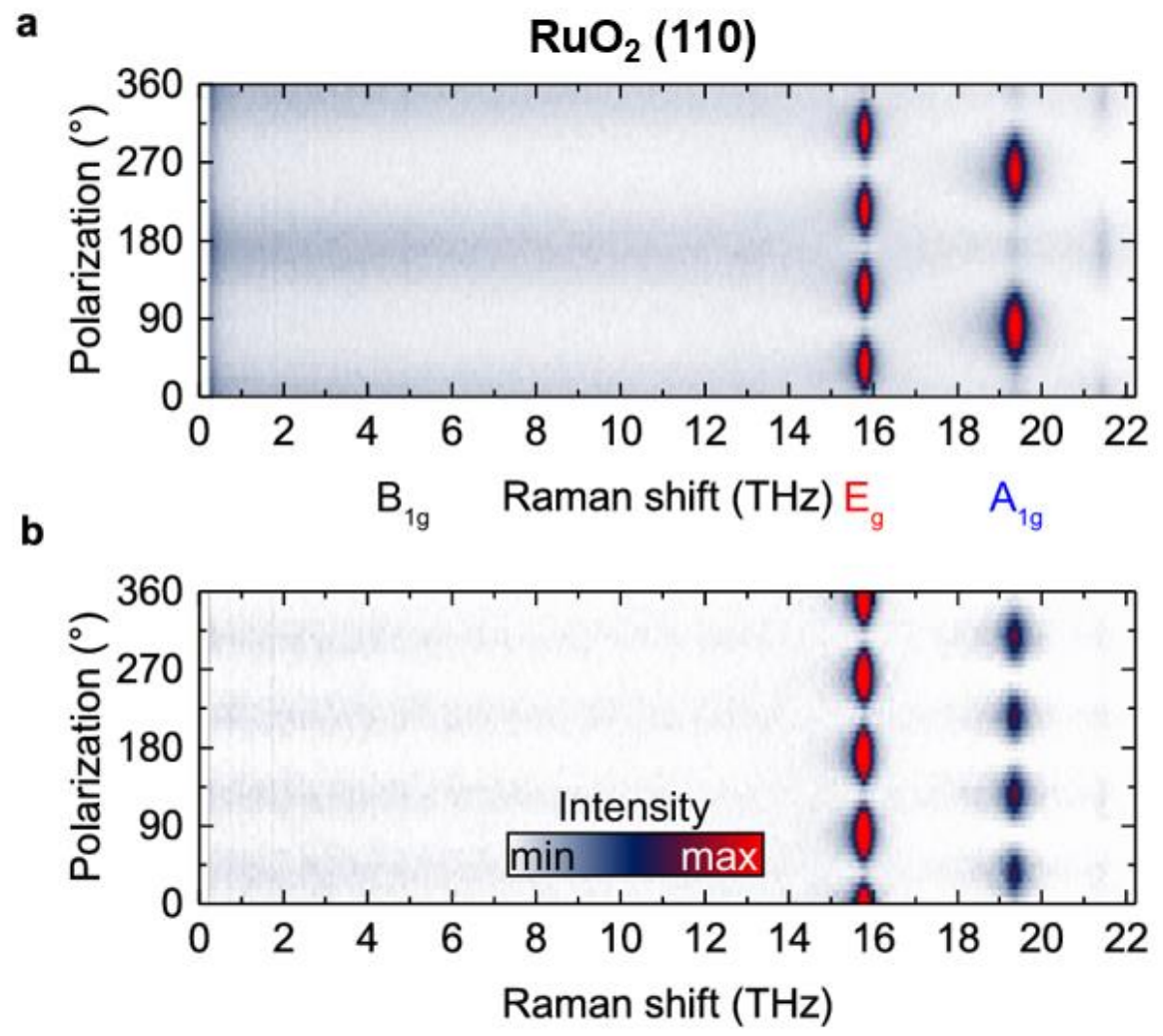

**Figure S17. Polarization-dependent Raman intensity for $RuO_2$ (110) single crystal a) $e_{\mathrm{in}}$//$e_{\mathrm{out}}$ and b) $e_{\mathrm{in}} \perp e_{\mathrm{out}}$ configurations.**

This explains the absence of the $B_{1g}$ mode in both Raman scattering and impulsive stimulated Raman scattering, as discussed in the main text, and is consistent with the Raman spectrum obtained from $RuO_2$ (110) single crystal samples (Fig. S17).

## Section S11. Probe-Polarization-Dependent FFT Sepctra of $B_{1g}$ Oscillation Obtained from Transient Magneto-Optic Signals

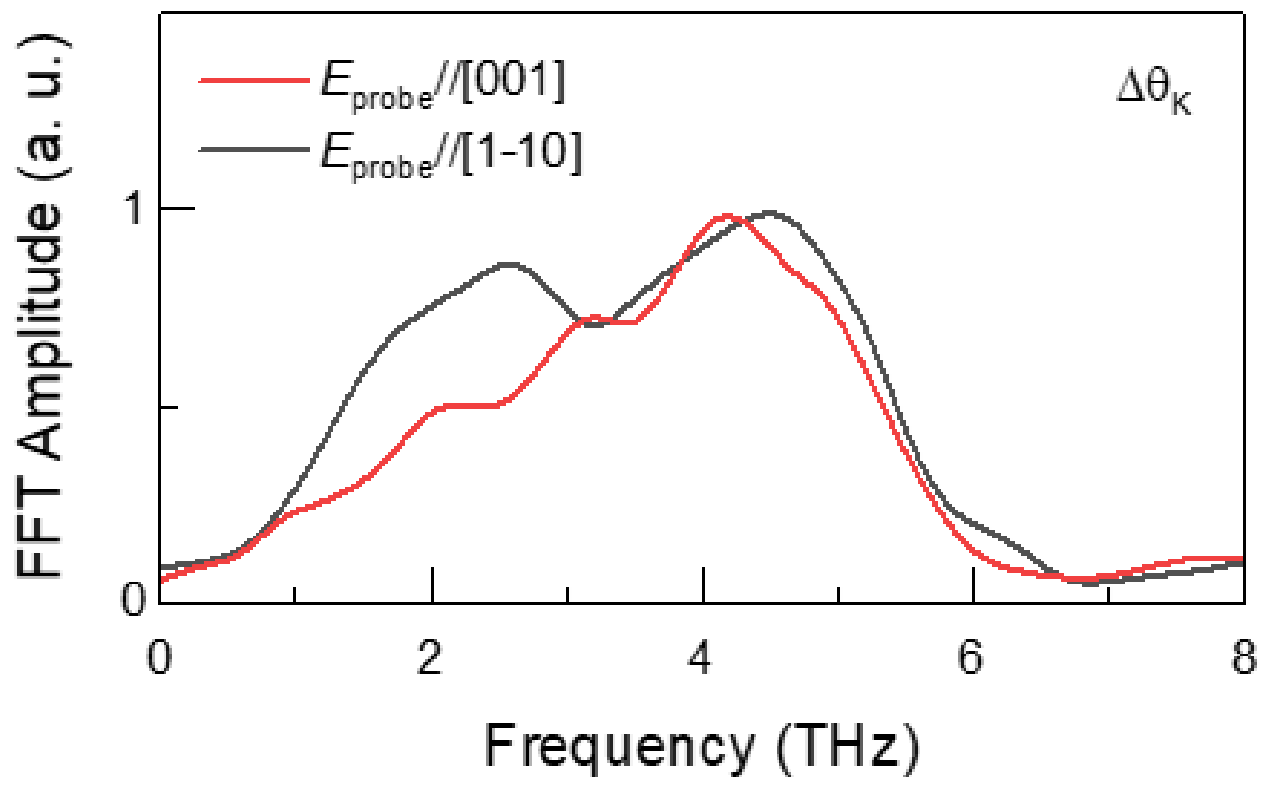

**Figure S18. FFT amplitude spectra obtained from the $B_{1g}$ coherent oscillation observed in the transient MOKE signal $\Delta\theta_K$ for the probe polarization along the [001] (red) and [1-10] (black).**

Based on the symmetry argument discussed in Section S10, the $B_{1g}$ mode observed in the MOKE signals should be induced indirectly, for example via spin–phonon coupling in the non-equilibrium state, as suggested in the main text. Since the MOKE signal in the polar configuration reflects the magnetization dynamics independent of the probe polarization[13], the $B_{1g}$ oscillation in the magnetization dynamics can be observed for any probe polarization. Figure S18 shows the FFT spectra of $B_{1g}$ oscillations obtained under photoexcitation for the probe polarization $E_{probe}$ parallel to [001] (red) and [1$\bar{1}$0] (black). Both the FFT spectra exhibit the identical center frequency around 4 THz with broad bandwidth, consistent to our expectation.

## Section S12. Temperature-dependent Raman Spectroscopy for $RuO_2$ Single Crystal

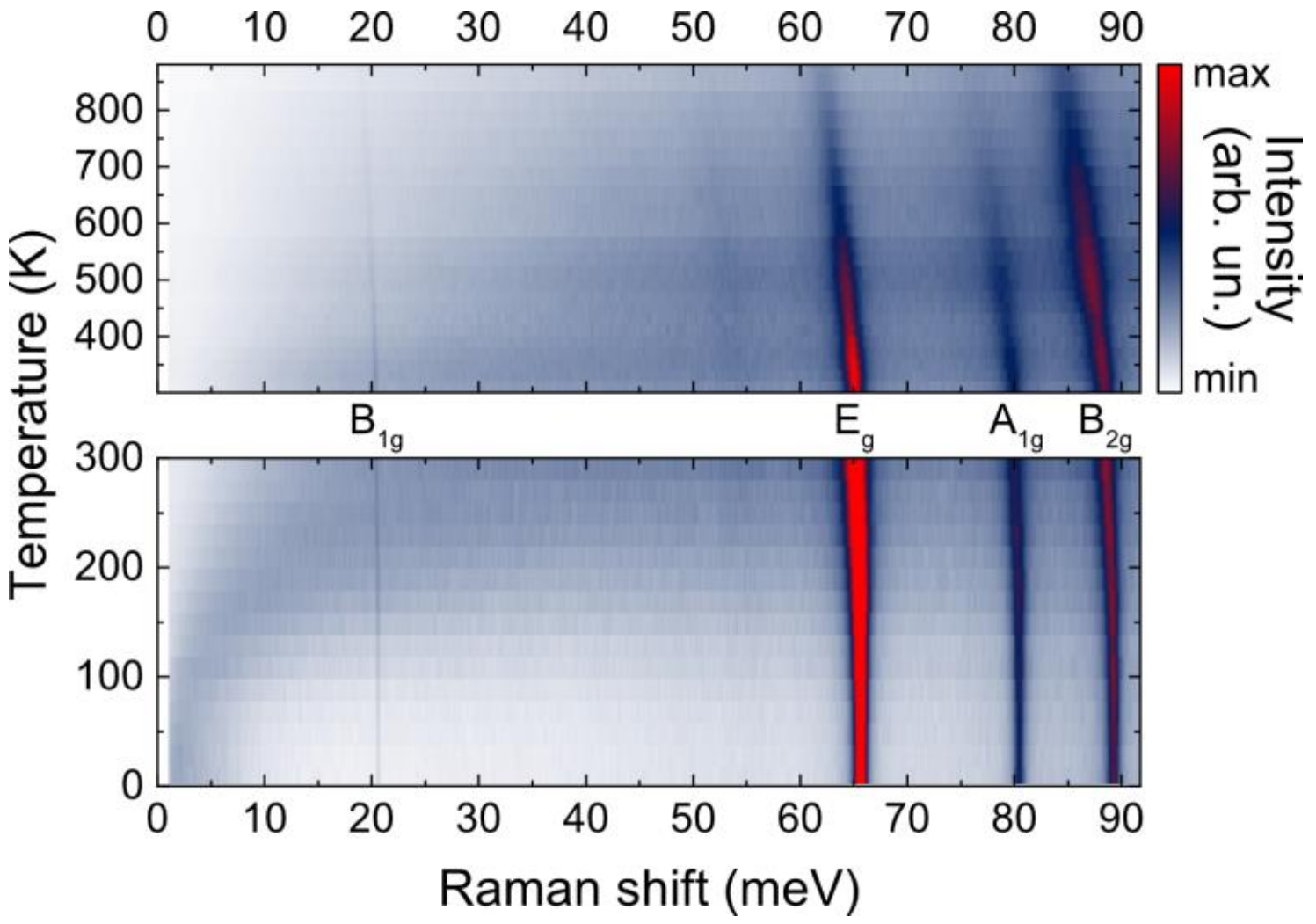

**Figure S19. Temperature-dependent unpolarized Raman spectrum for the (101) surface of a $RuO_2$ single crystal.**

To contrast the behavior between strained films and single crystals, we conducted temperature dependent Raman measurements on single crystalline $RuO_2$ in unpolarized condition, as shown in Fig. S19. Data was taken using an optical open-flow He cryostat (Oxford, $T$ = 5 K – 300 K) as well as a heating stage (Instec, $T$ = 300 K – 870 K). We find a continuous softening of phonon frequencies with increasing temperatures, while no fingerprint of a phase transition (structural or magnetic) is observed across this wide temperature range.

**Section S13. Comparison of Magnetization Dynamics Between 12 nm and 40 nm $RuO_2$ films**

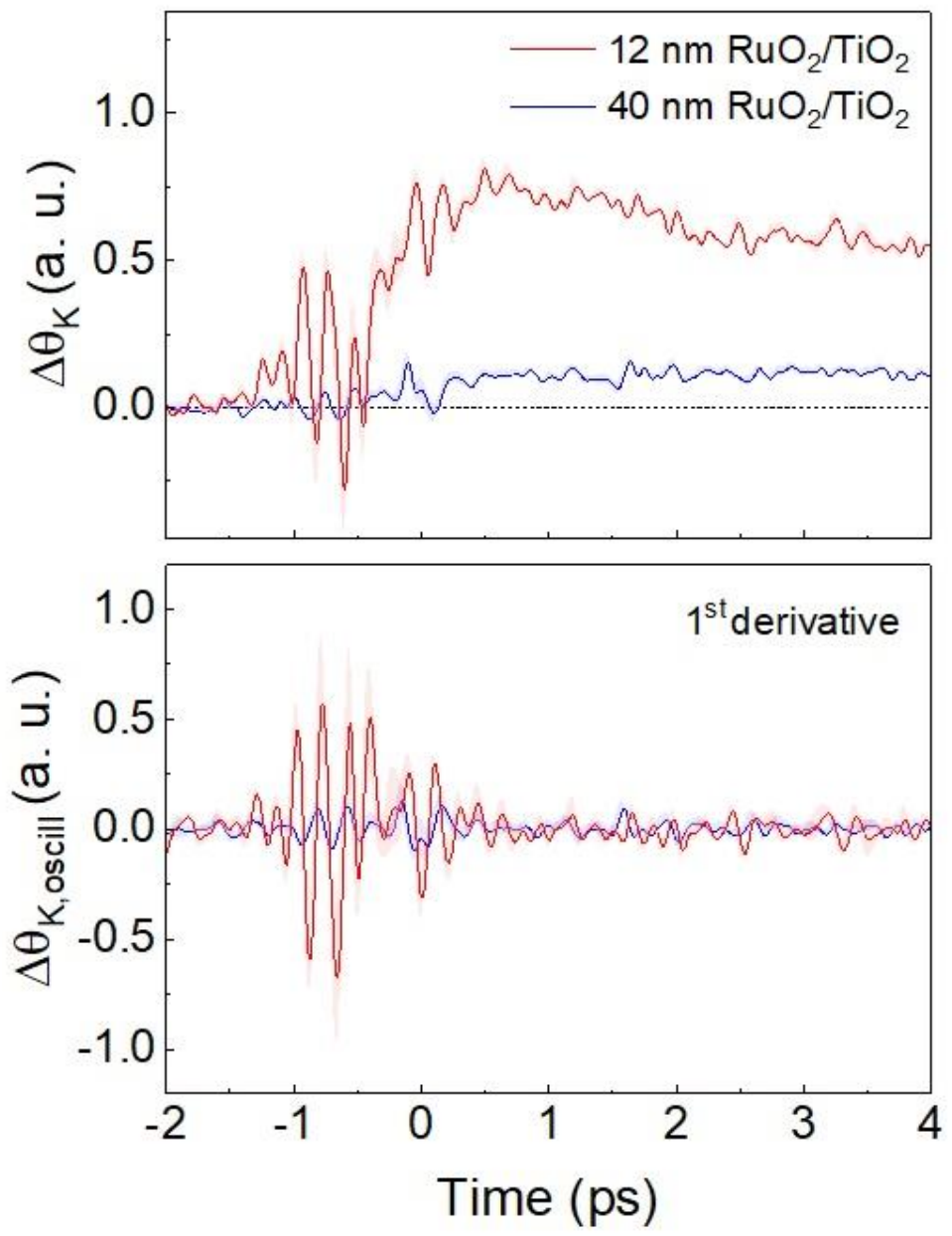

**Figure S20. Time-resolved magneto-optic Kerr effect results on 12 nm (red) and 40 nm (blue) $RuO_2$ films.**

To explore the role of local dislocations in $RuO_2$ films on magnetization dynamics, we conducted time-resolved magneto-optical Kerr effect (tr-MOKE) measurements on 12 nm and 40 nm $RuO_2/TiO_2$ (110) samples. We note that the 40 nm $RuO_2$ film exhibits bulk-like lattice constants and lacks observable local dislocations[14], while the 12 nm film contains a significant number of dislocations. Figure S20 exhibits the tr-MOKE results for the 12 nm (red) and 40 nm (blue) films under a magnetic field of 120 Oe (top panel), along with their first derivatives (bottom panel). The photo-induced change in $\theta_K$ ($\Delta\theta_K$) is significantly larger in the 12 nm film compared to the 40 nm film, suggesting that local dislocations hosting large strain are closely related to the emergence of magnetism in $RuO_2$. Furthermore, a clear THz oscillation is observed exclusively in the 12 nm film, well above the measurement error (shaded area), whereas no notable oscillation appears in the 40 nm film. This distinct difference in magnetization dynamics between the two films that have a distinct strain state can rule out experimental artifacts and highlight the critical role of strain in stabilizing the magnetic state in $RuO_2$.